\documentclass[journal,draftcls,onecolumn,romanappendices]{IEEEtran}
\usepackage{cite}
\usepackage{amsthm,amsmath,comment,bbm,amssymb,color,tikz,graphicx}
\usetikzlibrary{arrows}
\usepackage{subcaption}
\usepackage{array,comment}
\def\nbb{{\mathbf{b}}}
\def\nbc{{\mathbf{c}}}

\def\nbu{{\mathbf{u}}}

\def\nb0{{\mathbf{0}}}
\def\nb1{{\mathbf{1}}}

\def\ncalC{{\mathcal{C}}}

\def\ncalG{{\mathcal{G}}}

\def\ncalN{{\mathcal{N}}}

\def\ncalS{{\mathcal{S}}}

\def\ncalX{{\mathcal{X}}}

\def\nbbC{{\mathbb{C}}}

\def\nbbE{{\mathbb{E}}}

\def\nbbP{{\mathbb{P}}}

\def\nbbR{{\mathbb{R}}}

\def\nbbZ{{\mathbb{Z}}}

\newtheorem{lem}{Lemma}
\newtheorem{thm}{Theorem}
\newtheorem{ndef}{Definition}
\newtheorem{nrem}{Remark}

\newtheorem{cor}{Corollary}
\newtheorem{eg}{Example}
\newtheorem{assumption}{Assumption}

\begin{document}
  \title{Sensing using Coded Communications Signals}
  
  \author{Sundar Aditya, \IEEEmembership{Member, IEEE}, Onur Dizdar, \IEEEmembership{Member, IEEE}, Bruno Clerckx, \IEEEmembership{Fellow, IEEE}, and Xueru Li

  \thanks{This work was supported by Huawei Device Co., Ltd.}
  \thanks{S. Aditya, O. Dizdar and B. Clerckx are with the Dept.~of Electrical and Electronic Engineering, Imperial College London, London SW7 2AZ, U.K. (e-mail: \{s.aditya, o.dizdar, b.clerckx\}@imperial.ac.uk).}
  \thanks{X. Li is with Huawei Device Co., Ltd., Beijing, China (email: lixueru2@huawei.com) }
  }
 \maketitle

\begin{abstract}
A key challenge for a common waveform for Integrated Sensing and Communications (ISAC) -- widely seen as an attractive proposition to achieve high performance for both functionalities, while efficiently utilizing available resources -- lies in leveraging information-bearing channel-coded communications signals (c.c.s) for sensing. In this paper, we investigate the sensing performance of c.c.s in (multi-user) interference-limited operation, and show that it is limited by \emph{sidelobes} in the range-Doppler map, whose form depends on whether the c.c.s modulates a single-carrier or OFDM waveform. While uncoded communications signals -- comprising a block of $N$ i.i.d zero-mean symbols -- give rise to \emph{asymptotically} (i.e., as $N \rightarrow \infty$) \emph{zero sidelobes} due to the law of large numbers, it is not obvious that the same holds for c.c.s, as structured channel coding schemes (e.g., linear block codes) induce dependence across codeword symbols. In this paper, we show that c.c.s also give rise to asymptotically zero sidelobes -- for both single-carrier and OFDM waveforms -- by deriving upper bounds for the tail probabilities of the sidelobe magnitudes that decay as $\exp( - O(\mbox{code rate} \times \mbox{block length}))$. This implies that for any code rate, c.c.s are effective sensing signals that are robust to multi-user interference at \emph{sufficiently large} block lengths, with negligible difference in performance based on whether they modulate a single-carrier or OFDM waveform. We verify the latter implication through simulations, where we observe the sensing performance (characterized by the detection and false-alarm probabilities) of a QPSK-modulated c.c.s (code rate $= 120/1024$, block length $= 1024$ symbols) to match that of a comparable interference-free FMCW waveform even at high interference levels (signal-to-interference ratio of $-11{\rm dB}$), for both single-carrier and OFDM waveforms. % Consequently, in multi-user ISAC scenarios with a common waveform (either single-carrier or OFDM, but) largely comprising coded data symbols, sensing interference management essentially \emph{takes care of itself} and is much simpler than communications interference management.
%-- a highly desirable outcome in terms of maximizing communications spectral efficiency.
%it is sufficient to only focus on communications interference management, as sensing interference suppression is essentially realized \emph{for free}
\end{abstract}

\begin{IEEEkeywords}
Integrated Sensing and Communications (ISAC), Opportunistic sensing, Sensing in beyond 5G networks, Sensing interference management, Correlation properties of coded communications signals (c.c.s), Interference sidelobes
\end{IEEEkeywords}

\section{Introduction}
\label{sec:intro}
A combination of factors, such as (a) supporting emerging applications that require both communications and sensing functionalities (e.g., augmented/virtual reality, autonomous vehicles, etc.) in beyond 5G networks, and (b) the channel propagation characteristics at mmWave and THz frequency bands, have provided the impetus for the ongoing research efforts on integrated sensing and communications (ISAC). In this regard, the waveform used for each of the functionalities is a crucial aspect of system design. Broadly, there are two paradigms of waveform design for ISAC systems:
\begin{itemize}
    \item[a)] \emph{Separate waveforms}: Such an approach allows commercial stakeholders, who may have expertise in either communications or sensing but not both, to continue to independently design their hardware and systems, based on the desirable waveform properties for each functionality (e.g., FMCW for radar and OFDM for communications). The individual waveforms can either be transmitted orthogonally \cite{Alloulah_Huang_2019, Ma_et_al_2022} or superimposed \cite{Sahin_Arslan_2020}, provided the mutual interference is effectively mitigated.
    
    \item[b)] \emph{Common waveform}: The efficient utilization of valuable spectral, hardware and energy resources is the driving factor behind this approach \cite{Chiriyath_Paul_Bliss_tccn_2017}, with     \cite{Dokhanchi_et_al_radarconf_2018, Ozkaptan_et_al_2018, Ozkaptan_et_al_infocom_2020} being good examples that focus on optimizing the allocation of resources (e.g., power, subcarriers etc.) between communications and sensing functionalities for a common OFDM waveform. In contrast to these, a special case of a common waveform involves sensing using \emph{only} the reference signals of a primarily communications waveform, such as IEEE 802.11ad frames \cite{Kumari_Heath_tvt_2018, Grossi_etal_tsp_2018, Ajorloo2019OnTF, Grossi_etal_twc_2020}. While such a solution has the advantage of not requiring extensive standardization efforts, it is sub-optimal as, in principle, even the communications payload/signal within a common waveform can be used for sensing in a monostatic configuration, since it is known apriori at the transmitting node. 
\end{itemize}
In this paper, we explore the sensing potential of (channel) coded communications signals (c.c.s) -- i.e., the communications payload -- when embedded in a common ISAC waveform. A key question w.r.t a common ISAC waveform is whether to adopt a single- or multi-carrier waveform, with most discussion of the benefits of one over the other focusing on issues like signal processing complexity (both communications and sensing), easy integration with existing standards, PAPR etc. \cite{Wild_Braun_Vishwanathan_2021, Kai_et_al_2022}. Our focus on the single- versus multi-carrier question is restricted to the impact, if any, of channel coding on the sensing performance of each waveform, especially in interference-limited environments.

The principle of using known data symbols for sensing has been explored for a variety of candidate waveforms, such as single-carrier \cite{Zeng_Ma_Sun_2020}, OFDM \cite{Sit_et_al_spacomm_2011, Barneto_et_al_tmt_2019, Guan_et_al_2020, Dokhanchi_et_al_radarconf_2018, Liyanaarachchi_et_al_2020}, and OTFS (Orthogonal Time Frequency and Space)\cite{Gaudio_Caire_2020}. Among these, \cite{Sit_et_al_spacomm_2011, Dokhanchi_et_al_radarconf_2018} and \cite{Gaudio_Caire_2020} consider uncoded communications signals -- unrepresentative of typical data payloads that are subject to channel coding -- while the assumptions around channel coding in \cite{Zeng_Ma_Sun_2020, Barneto_et_al_tmt_2019} and \cite{Guan_et_al_2020} are unclear. In particular, the latter two references are experimental studies exploring the sensing potential of 5G NR signals, and hence, even if channel coding was implied, its impact on the sensing performance was not investigated. Finally, none of the above references address the impact of interference on the sensing performance of c.c.s.

\begin{figure}
    \centering
\scalebox{0.9}{\begin{tikzpicture}
       \draw[thick,black,-] (-2.5,0) -- (0,0) -- (0,1) -- (-2.5,1) -- (-2.5,0) node at (0.75-2,0.5){\small TX$/$Radar~1};
       \draw[thick,black,-] (2-2,0.5) -- (2.3-2,0.5) -- (2.3-2,-1) -- (2-2,-1.3) -- (2.6-2,-1.3) -- (2.3-2,-1) node at (2.3-2,-2){\vdots};
       \draw[thick,black,-] (-2.5,-5) -- (0,-5) -- (0,-4) -- (-2.5,-4) -- (-2.5,-5) node at (0.75-2,-4.5){\small TX$/$Radar~$M_r$};
       \draw[thick,black,-] (2-2,-4.5) -- (2.3-2,-4.5) -- (2.3-2,-3) -- (2-2,-2.7) -- (2.6-2,-2.7) -- (2.3-2,-3);
       \draw[thick,black,-] (-5,0) -- (-4,0) -- (-4,1) -- (-5,1) -- (-5,0) node at (-4.5,0.5){\small RX~1};
       \draw[thick,black,-] (-4,0.5) -- (-3.7,0.5) -- (-3.7,-1) -- (-4,-1.3) -- (-3.4,-1.3) -- (-3.7,-1) node at (-3.7,-2){\vdots};
       \draw[thick,black,-] (-5,-5) -- (-4,-5) -- (-4,-4) -- (-5,-4) -- (-5,-5) node at (-4.5,-4.5){\small RX~$M_r$};
       \draw[thick,black,-] (-4,-4.5) -- (-3.7,-4.5) -- (-3.7,-3) -- (-3.4,-2.7) -- (-4,-2.7) -- (-3.7,-3);
       \draw[thick,magenta,->] (1.6-1.8,-1.3) -- (-3,-1.3) node at (-1.5,-1){Comm.~link 1};
       \draw[thick,magenta,->] (1.6-1.8,-2.7) -- (-3,-2.7) node at (-1.5,-3){Comm.~link $M_r$};
       \draw[thick,magenta,dashed,->] (1.6-1.8,-1.4) -- (-3,-2.6);
       \draw[thick,magenta,dashed,->] (1.6-1.8,-2.6) -- (-3,-1.4) node at (1-2,-2){Comm.~intf.};
       \draw[black,fill=black] (2.5,-1.3) circle (0.1) node at (2.5,-0.85){\small Target 1} node at (2.5,-2){\vdots};
       \draw[black,fill=black] (2.5,-2.7) circle (0.1) node at (2.5,-3.15){\small Target $N_{\rm tar}$};
       \draw[thick,red,<->] (0.7,-1.3) -- (2.4,-1.3);
       \draw[thick,red,<->] (0.7,-1.35) -- (2.35,-2.65);
       \draw[thick,blue,<->] (0.7,-2.7) -- (2.4,-2.7) ;
       \draw[thick,blue,<->] (0.7,-2.55) -- (2.35,-1.4) ;
      \draw[thick,red,dashed,->] (0.8,-1.4) .. controls (1,-2) .. (0.8,-2.6);
      \draw[thick,red,dashed,->] (0.8,-1.4) -- (2.5,-1.4) -- (0.8,-2.6);
      \draw[thick,blue,dashed,->] (0.6,-2.6) .. controls (0.8,-2) .. (0.6,-1.4);
      \draw[thick,blue,dashed,->] (0.6,-2.6) -- (2.5,-2.6) -- (0.9,-1.4);
      \draw[thick, red, <->] (1,2) -- (2.5,2);
      \draw[thick, blue, <->] (1,1.8) -- (2.5,1.8);
      \draw[thick,black,-] (0.75,2.25) -- (5.75,2.25) -- (5.75,1.15) -- (0.75,1.15) -- (0.75,2.25);
      \node[black] at (4.1,1.9){\small Desired radar returns};
      \draw[thick, dashed, red, ->] (1,1.5) -- (2.5,1.5);
      \draw[thick, dashed, blue, ->] (1,1.3) -- (2.5,1.3);
      \node[black] at (3.5,1.4){\small Sensing intf.};
      \draw[very thick,black,-] (3.5,1) -- (3.5,-5) (3.6,1) -- (3.6,-5) node at (-0.5,-5.5){(a)} node at (7.5,-5.5){(b)};
      \draw[thick,black,-] (6,0.5) -- (6,-4.5) (9,0.5) -- (9,-4.5);
      \draw[thick,gray,fill=yellow] (7.4,0.5) -- (7.6,0.5) -- (7.6,0) -- (7.4,0) -- (7.4,0.5) (7.4,-1) -- (7.6,-1) -- (7.6,-1.5) -- (7.4,-1.5) -- (7.4,-1) (7.4,-2.5) -- (7.6,-2.5) -- (7.6,-3) -- (7.4,-3) -- (7.4,-2.5) (7.4,-4) -- (7.6,-4) -- (7.6,-4.5) -- (7.4,-4.5) -- (7.4,-4);
      \draw[thick,black,-] (6.5,-3.8) -- (7,-3.8) -- (7,-4.3) -- (6.5,-4.3) -- (6.5,-3.8) node at (6.75,-4.75){\small TX/Radar 1};
      \draw[thick,black,-] (8,0) -- (8.5,0) -- (8.5,-0.5) -- (8,-0.5) -- (8,0) node at (8.25,0.3){\small TX/Radar 2};
      \draw[thick,black,fill=black] (7.5,-2) circle (0.1) node at (8.1,-2){\small Target};
      \draw[thick,red,<->] (6.75,-3.7) -- (7.45,-2.1);
      \draw[thick,blue,<->] (7.6,-1.9) -- (8.25,-0.6);
      \draw[thick,red,dashed,->] (6.5,-3.7) -- (8.1,-0.6);
      \draw[thick,blue,dashed,->] (8.4,-0.6) -- (6.9,-3.7);
      \draw[thick,black,-] (4.5,-3.8) -- (5,-3.8) -- (5,-4.3) -- (4.5,-4.3) -- (4.5,-3.8) node at (4.75,-4.75){\small RX 1} node at(4.75,-5.2){\small (RSU)};
      \draw[thick,black,-] (10,0) -- (10.5,0) -- (10.5,-0.5) -- (10,-0.5) -- (10,0) node at (10.25,0.75){\small RX 2} node at (10.25,0.25){\small (RSU)};
      \draw[thick,magenta,->] (6.4,-4) -- (5.1,-4) node at (5.4,-3.6){\small Comm.~link~1};
     \draw[thick,magenta,->] (8.55,-0.25) -- (9.9,-0.25) node at (9.7,-0.55){\small Comm.~link~2};
    \end{tikzpicture}
    }
    \caption{(a) Using a common transmit waveform, TX~$i$ communicates with RX~$i$ ($i=1,\cdots,M_r)$, while all TXs/radars simultaneously sense a common target scene non-cooperatively. This gives rise to distinct communications and sensing interference, where the latter often dominates the former due to a combination of link geometry and/or greater pathloss experienced by the desired radar returns; e.g., in (b) a vehicular network that can be interpreted as a special case of (a), corresponding to $M_r = 2$ and $N_{\rm tar} = 1$.}
    \label{fig:setup}
\end{figure}
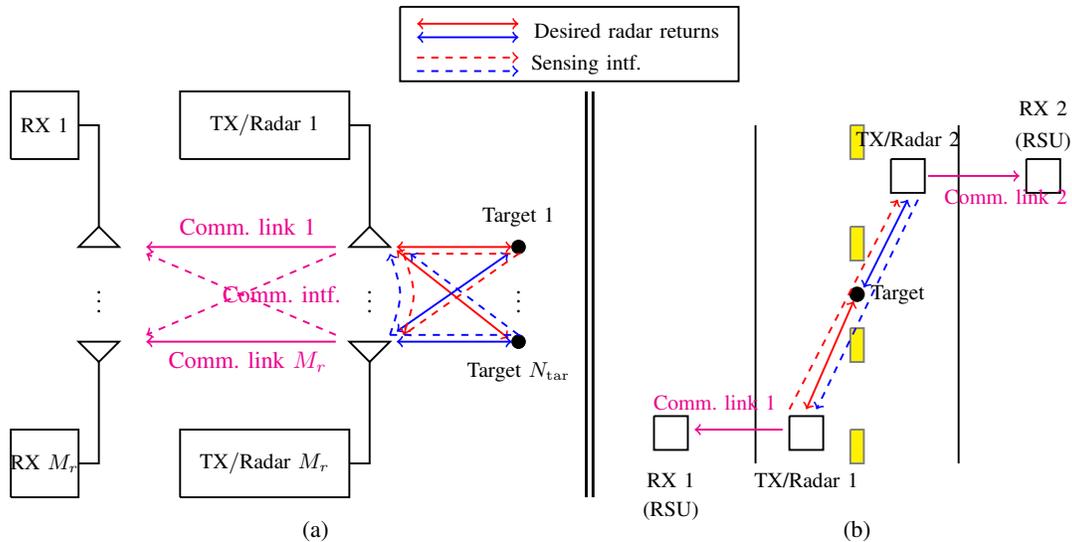
To motivate the discussion on the sensing potential of c.c.s in interference-limited operating environments, consider the scenario in Fig.~\ref{fig:setup}a, where TX~$i$ communicates with RX~$i$ ($i=1,\cdots,M_r$), while all the TXs also act as monostatic radars to simultaneously sense a scene without cooperation, using a common ISAC waveform. Such a setup gives rise to \emph{multi-user} communications and sensing interference (highlighted in Fig.~\ref{fig:setup}a), where the former is the interference experienced at RX~$i$ due to transmissions of TX~$j(\neq i)$, and the latter is the interference at  radar receiver~$i$ (colocated with TX~$i$) due to the transmissions of TX/radar~$j(\neq i)$\footnote{Note that our definition of multi-user sensing interference is distinct from adversarial interference, like jamming, which we do not consider in this paper. For brevity, we drop the `multi-user' prefix when discussing sensing interference from here on.}. In general, the two types of interference are distinct in their profiles, and we are especially interested in cases where the sensing interference \emph{dominates} the communications interference; an example of such a scenario could occur in a vehicular network, as depicted in Fig.~\ref{fig:setup}b, where a pair of vehicular TXs/radars (i.e., $M_r = 2$) are facing each other on opposite lanes and sensing the road ahead of them, while also communicating with their RXs, which may be the nearest road side units (RSUs) located on their respective sidewalks. The geometry of this scene is such that even with large antenna arrays, beamforming may be able to effectively suppress communications -- but not sensing -- interference, as even with narrow beams, there could still be significant signal leakage along the direct (line-of-sight) path linking the two TXs. Geometry apart, the $R^{-4}$ attenuation of the desired radar returns -- in contrast to the $R^{-2}$ attenuation for the interference signal along the direct path between two radars -- also makes it more likely that sensing interference dominates communications interference in other ISAC use cases. Hence, at first glance, it is tempting to allocate the available communications resources (e.g., time and frequency bands) orthogonally across TXs to mitigate the sensing interference, at the expense of communications spectral efficiency (measured in bits/s/Hz). However, if c.c.s happen to be \emph{good} sensing signals -- meaning, achieving high sensing performance, while being robust to sensing interference -- then, the communications spectral efficiency in Fig.~\ref{fig:setup} can be significantly enhanced through a reuse factor of one. 

To provide intuition for why communications signals might be good sensing signals, consider a collection of $M_r$ \emph{uncoded} communications signals -- $\{v^{(i)}[n]: n=0,\cdots,N-1;~ i = 1,\cdots,M_r\}$ -- comprising i.i.d zero mean, unit-energy symbols (e.g., QPSK). From the law of large numbers (LLN), it can easily be shown that these signals have \emph{asymptotically favourable correlation properties} (i.e., auto-correlation function tending to the $\delta$-function, and mutual cross-correlation function tending to the zero function, as $N \rightarrow \infty$), which -- as shown in Section~\ref{subsubsec:sensing_sc} -- makes them good sensing signals, when embedded in a single-carrier waveform. Similarly, the IDFT of $v^{(q)}[n]/v^{(i)}[n]~(q \neq i)$ identically tends to 0, as $N \rightarrow \infty$, which -- as shown in Section~\ref{subsubsec:sensing_ofdm} -- makes the collection of signals $\{v^{(i)}[n]: i=1,\cdots,M_r\}$ robust to sensing interference, when embedded in an OFDM waveform. However, structured state-of-the-art channel coding schemes (e.g., linear block codes, such as Polar and LDPC codes) induce statistical dependence across symbols, whose impact on the above functions of c.c.s has not yet been characterized, to the best of our knowledge. Hence, we tackle these questions in this paper, with a view to determine if (a) c.c.s can indeed achieve high sensing performance, while being robust to sensing interference, and (b) whether the nature of the waveform (i.e., single- v/s multi-carrier) influences the answer to (a). Our contributions are as follows:
\begin{itemize}
    \item[1.] \textbf{Single-Carrier waveform}: 

    \begin{itemize}
    \item We derive an upper bound for the tail probability of (the non-zero lags of) the auto-correlation function of c.c.s that decays as $\exp(-O(rN))$, where $r$ is the code rate and $N$ the block length (Theorem~\ref{thm:main}). As a result, c.c.s have an asymptotically ideal auto-correlation function that converges in distribution to (a suitably scaled) $\delta$-function for large block lengths (Corollary~\ref{cor:asymp}), which makes them good sensing signals, when embedded in a single-carrier waveform. These results makes mild assumptions regarding c.c.s and importantly, do not depend on the code structure.
    
    \item For a pair of c.c.s with rates $r_i$ and $r_q$, generated from independent message signals, we derive an upper bound for the tail probability of their cross-correlation function that decays as $\exp(-O(\max (r_i,r_q) N))$ (Corollary~\ref{cor:main}). As a result, c.c.s generated by independent message signals have a cross-correlation function that converges in distribution to the zero function for large block lengths (Corollary~\ref{cor:asymp_crosscorr}), which makes them robust to sensing interference, when embedded in a single-carrier waveform.
    
    \item  For the above setup, but with linear codes, we derive lower bounds for the tail probabilities of the auto- and cross-correlation functions of c.c.s that also decay as $\exp(-O(rN))$ (Theorem~\ref{thm:lb} and Corollary~\ref{cor:lb}, respectively). These results, along with the previous bullet points, imply that for linear codes, a faster rate of decay for these functions, as a function of $N$, is not possible.
    \end{itemize}
    
\item[2.] \textbf{OFDM waveform}: 
    \begin{itemize}
    \item For a pair of c.c.s -- $s^{(i)}[n]$ and $s^{(q)}[n]$, with rates $r_i$ and $r_q$, respectively -- generated from independent message signals, we derive an upper bound for the tail probability of the IDFT of $s^{(q)}[n]/s^{(i)}[n]$ that decays as $\exp(-O(\min(r_i,r_q) N)$ (Corollary~\ref{cor:idft_sidelobe}). As a result, the above IDFT converges in distribution to the zero function for large block lengths (Corollary~\ref{cor:asymp_idft}), which makes c.c.s robust to sensing interference, when embedded in an OFDM waveform.
        
    \item For the same setup as the previous bullet point, but with linear codes, we derive a lower bound for the tail probability of the IDFT of $s^{(q)}[n]/s^{(i)}[n]$, that also decays as $\exp(-O(rN))$ (Corollary~\ref{cor:ofdm_lb}). The latter, along with the previous bullet point, implies that for linear codes, a faster rate of decay for the IDFT, as a function of $N$, is not possible.
   \end{itemize}
\item[3.] \textbf{Implications}:
    \begin{itemize}
    \item The results under points 1 and 2 above suggest that for any code rate, c.c.s can be effective sensing signals that are robust to sensing interference at \emph{sufficiently large} block lengths, with negligible difference in performance based on whether they modulate a single-carrier or OFDM waveform. We verify this implication through simulations, where we observe that the sensing performance (characterized by the detection and false-alarm probabilities) of a QPSK-modulated c.c.s (code rate $= 120/1024$, block length $= 1024$ symbols) matches that of a comparable interference-free FMCW waveform even at a signal-to-interference ratio of $-11{\rm dB}$, for both single-carrier and OFDM waveforms.
    
    \item Thus, a common ISAC waveform (either single- or multi-carrier, but) largely comprising coded data symbols is an effective sensing signal at large block lengths that can also achieve high communications spectral efficiency. Moreover, in multi-user ISAC scenarios with such a waveform, sensing interference management essentially \emph{takes care of itself} for monostatic radars, and is relatively simpler than communications interference management, which is favourable for the evolution of existing wireless broadband networks to support sensing applications.
    \end{itemize}
\end{itemize}

%  We verify the latter implication through simulations, where we observe . {\color{magenta}Consequently, in multi-user ISAC scenarios with a common waveform (either single-carrier or OFDM, but) largely comprising coded data symbols, it is sufficient to only focus on communications interference management, as sensing interference suppression is essentially realized \emph{for free} -- a highly desirable outcome in terms of maximizing communications spectral efficiency.}
This paper consists of five sections. In our system model in Section~\ref{sec:sys_model}, we identify functions of c.c.s -- depending on whether they modulate a single-carrier or OFDM waveform -- whose tail distribution needs to decay rapidly, for c.c.s to be good sensing signals that are robust to sensing interference. In Section~\ref{sec:corr}, we characterize the tail probabilities of these functions and show that they decay rapidly at large block lengths. We verify the effectiveness of c.c.s as sensing signals in interference-limited environments through simulations in Section~\ref{sec:sim_res}, where we see that the sensing performance of c.c.s -- for both single-carrier and OFDM waveforms -- is at par with the (interference-free) FMCW waveform, even at a signal-to-interference ratio (SIR) of $-11{\rm dB}$. Finally, we conclude this paper in Section~\ref{sec:conc_rems} with some remarks on the implications of our results.

\subsection*{Notation:}Vectors are represented by lower case bold letters; $\ncalC\ncalN(0,\sigma^2)$ denotes the circularly symmetric complex Gaussian distribution with mean zero and variance $\sigma^2$; $\delta[\cdot]$ and $\mathbbm{1}(.)$ denote the discrete delta and the indicator functions, respectively; $\nbbP(\cdot)$ denotes probability; $\nbbE[\cdot]$ denotes the expectation operator, with $\nbbE_X[\cdot]$ explicitly denoting the expected value w.r.t the random variable, $X$; $(\cdot)^*$ denotes complex conjugation, and $\Re\{\cdot\}$ the real part. We assume that all discrete-time signals are supported on $\nbbZ$, the set of integers, with the understanding that for a signal, $s[n]~(n=0,\cdots,N-1)$, with finite support, $s[n]=0$ for $n \notin \{0,\cdots,N-1\}$. Finally, we introduce a few definitions that will be used throughout the paper:

% $a_i$ denotes the $i$-th element of vector $\nba$;
% , and $\xrightarrow{w.p.1}$ convergence with probability one (almost sure convergence)
%, $Q_1(.,.)$ denotes the Marcum Q-function, and $\xrightarrow{w.p.1}$ denotes convergence with probability one (almost sure convergence). Finally, the auto- and cross-correlation functions are defined as follows:

\begin{ndef}[Auto-correlation function]
\label{def:autocorr}
 For a discrete-time signal, $s[n]~(n=0,\cdots,N-1)$, its (aperiodic) auto-correlation function at lag $l \in \nbbZ$, denoted by $\chi(l;s[\cdot])$, is given by:
\begin{IEEEeqnarray}{rCl}
\label{eq:chi_s}
    \chi(l;s[\cdot]) & := &\frac{1}{N} \sum_{n = -\infty}^{\infty} s[n] s^*[n-l].
\end{IEEEeqnarray}
\end{ndef}

\begin{ndef}[Cross-correlation function]
\label{def:crosscorr}
For a pair of signals, $s_1[n]$ and $s_2[n]~(n=0,\cdots,N-1)$, their (aperiodic) cross-correlation function at lag $l\in \nbbZ$, denoted by $\varrho(l;s_1[\cdot],s_2[\cdot])$, is given by:
\begin{IEEEeqnarray}{rCl}
\label{eq:varrho}
    \varrho(l; s_1[\cdot],s_2[\cdot]) & := & \frac{1}{N} \sum_{n=-\infty}^{\infty} s_1[n] s_2^*[n-l].
\end{IEEEeqnarray}
\end{ndef}
Clearly, by definition, $\chi(l;s[\cdot]) = \varrho(l;s_1[\cdot],s_2[\cdot]) = 0$ for $|l| \geq N$. Hence, in this paper, we focus on the behavior of $\chi(l;s[\cdot])$ and $\varrho(l;s_1[\cdot],s_2[\cdot])$ for $|l| \ll N$, for large $N$. Alternately, for $N$-length signals, periodic auto- and cross-correlation functions can be defined by replacing the index $n-l$ with $(n-l) \mod N$ in (\ref{eq:chi_s}) and (\ref{eq:varrho}), respectively. Our results in this paper are valid for these functions, as well.

\begin{ndef}[Convergence in Distribution]
\label{def:conv_d}
A sequence of random variables, $X_1,\cdots,X_n$ converges in distribution to a random variable, $X$ -- denoted by $\displaystyle\lim\limits_{n \rightarrow \infty} X_n \xrightarrow{d} X$ -- if the following condition is satisfied:
\begin{IEEEeqnarray}{rCl}
    \lim_{n\rightarrow \infty} \nbbP(X_n > u) & = & \nbbP(X > u),
\end{IEEEeqnarray}
for all $u$ where the tail distribution, $\nbbP(X > u)$, is continuous.
\end{ndef}

\section{System Model}
\label{sec:sys_model}
\begin{figure}
 \centering
\scalebox{0.9}{\begin{tikzpicture}
   \draw[thick,-] (0,3) -- (8,3);
   \draw[thick,-] (8.15,2.75)+(0.05,0.5) -- +(-0.1,-0.1) -- +(0.1,0.1) -- +(0,-0.5);
   \draw[thick,-] (8.3,3) -- (15,3) -- (15, 2.5) -- (8.3,2.5);
   \draw[thick,-] (8,2.5) -- (0,2.5) -- (0,3);
  \draw[-] (0.5,3) -- (0.5,2.5) node at (0.25,2.75){\small $0$} node at (0.75,2.75){\small $1$};
  \draw[-] (1,3) -- (1,2.5) node at (1.5,2.75){$\cdots$};
  \draw[-] (4.5,3) -- (4.5,2.5);
  \draw[<-] (4.75,2.75) -- (4.75,2.2) node at (4.75,2.1){\small{$N-1$}};
  \draw[-] (5,4.2) -- (5,2.2);
  \draw[-] (5.5,3) -- (5.5,2.5);
  \draw[-] (6,3) -- (6,2.5) node at (6.5,2.75){$\cdots$};
  \draw[-] (10,4.2) -- (10,2.2);
  \draw[-] (10.5,3) -- (10.5,2.5);
  \draw[-] (11,3) -- (11,2.5) node at (11.5,2.75){$\cdots$};
  \draw[-] (14.5,3) -- (14.5,2.5);
%  \draw[<-] (14.75,2.75) -- (14.75,2.2) node at (14.75,2.2){\small{$MN-1$}};
  \draw[<->] (0,4) -- (5,4) node at (2.5,4.2){Slow time} node at (2.5,2){$x^{(i)}[n,1]$};
  \draw[<->] (0,5) -- (15,5) node at (7.5,5.2){Coherent Processing Interval (CPI)};
  \draw[-] (0,3.1) -- (0,3.3) (0.5,3.1) -- (0.5,3.3) node at (12.5,2){$x^{(i)}[n,M]$};
  \draw[->] (-0.5,3.2) -- (0,3.2);
  \draw[<-] (0.5,3.2) -- (1,3.2);
  \draw[->] (-0.5,3.5) -- (0,3.5) -- (0.25,3.2) node at (-1,3.5){Fast time};
  \node at (7.5,1.5){(a) c.c.s. (transmitted by TX/radar $i$)};
  \draw[thick,-] (0,0) -- (8,0) (8.3,0) -- (15,0) -- (15, -0.5) -- (8.3,-0.5) (8,-0.5)-- (0,-0.5) -- (0,0);% node at (-1,-0.25){$s_c[n]$};
  \draw[thick,-] (8.15,-0.25)+(0.05,0.5) -- +(-0.1,-0.1) -- +(0.1,0.1) -- +(0,-0.5);
  \draw[->] (12,-2)--(14,-2) node at (14.5,-2){Time};
  \draw[-] (0.5,0) -- (0.5,-0.5);
  \draw[-] (1,0) -- (1,-0.5) node at (1.5,-0.25){$\cdots$};
  \draw[-] (5,0.5) -- (5,-1);
  \draw[-] (5.5,0) -- (5.5,-0.5);
  \draw[-] (6,0) -- (6,-0.5) node at (6.5,-0.25){$\cdots$};
  \draw[-] (10,0.5) -- (10,-1);
  \draw[-] (10.5,0) -- (10.5,-0.5);
  \draw[-] (14.5,0) -- (14.5,-0.5);
  \draw[-] (11,0) -- (11,-0.5) node at (11.5,-0.25){$\cdots$} node at (2.5,-1){$x_r[n]$} node at (12.5,-1){$x_r[n]$};
  \node at (7.5,-1.5){(b) FMCW};
 \end{tikzpicture}
 }
 \caption{Sensing signal structure in time-domain for range-Doppler estimation.}
 \label{fig:fmcw_vs_ccs}
\end{figure}
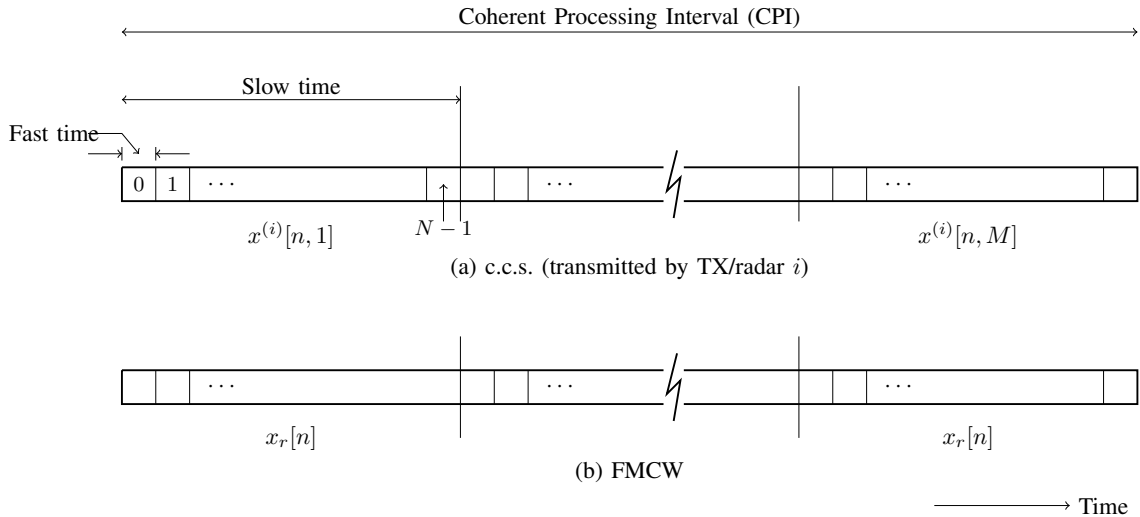

Consider the scenario in Fig.~\ref{fig:setup}a, where TX~$i$ communicates with RX~$i$ ($=1,\cdots,M_r$), while all the TXs also act as monostatic radars to simultaneously sense a scene comprising $N_{\rm tar}$ targets, without cooperation. Let $\ncalX^{(i)}_{M,N} := \{x^{(i)}[n,m] \in \nbbC : m = 1,\cdots,M;~n = 0,\cdots,N-1\}$ denote the (discrete) time-domain representation of the common ISAC waveform transmitted by the $i$-th TX/radar, which can be interpreted as a collection of $M$ signals, each of length $N$ -- i.e., $\{x^{(i)}[n,1], \cdots, x^{(i)}[n,M]: n=0,\cdots,N-1 \}$ -- where $x^{(i)}[n, 1]$ is transmitted first, followed by $x^{(i)}[n, 2]$, and so on, as shown in Fig.~\ref{fig:fmcw_vs_ccs}a\footnote{Typically, in purely radar applications, $x^{(i)}[n, m]$ is identical $\forall m \in \{1,\cdots,M\}$ (e.g., Fig.~\ref{fig:fmcw_vs_ccs}b). However, this is not a strict requirement.}. The indices, $n$ and $m$, represent two different time scales, with the \emph{fast time} ($n$) capturing variations in the round-trip delays associated with the target ranges, while the \emph{slow time} ($m$) is better suited to capture variations in the Doppler frequencies associated with moving targets. In general, for fixed $m$, each signal $x^{(i)}[n,m]$ is a function of a c.c.s, $s^{(i)}[n,m]$, whose modeling is described below.

\subsection{c.c.s Model}
\label{subsec:ccs_model}
\begin{figure}
    \centering
\scalebox{0.9}{\begin{tikzpicture}
     \draw[->] (-0.5,0) -- (1,0) node at (0.2,0.25){$\nbb_m^{(i)}$};
     \draw[thick,-] (1,0.5) -- (2,0.5) -- (2,-0.5) -- (1,-0.5) -- (1,0.5) node at (1.5,0){$\ncalG(\cdot)$} node at (1.5,0.8){Generator};
     \draw[->] (2,0) -- (3.5,0) node at (2.75,0.25){$\nbc_m^{(i)}$};
    \draw[thick,-] (3.5,0.5) -- (4.5,0.5) -- (4.5,-0.5) -- (3.5,-0.5) -- (3.5,0.5) node at (4,0){$\pi(\cdot)$} node at (4,0.8){Interleaver};
     \draw[->] (4.5,0) -- (6,0) node at (5.5,0.25){$\tilde{\nbc}_{m}^{(i)}$};
     \draw[thick,-] (6,1) -- (8.2,1) -- (8.2,-1) -- (6,-1) -- (6,1) node at (7.1,0){\begin{tabular}{c} $m_{{\rm s}_i}$ \\  bits/symb \end{tabular}};
     \draw[->] (8.2,0.5) -- (9,0.5) node at (12,0.75){$s^{(i)}[0,m], s^{(i)}[1,m], \cdots, s^{(i)}[N-1,m]$};
     \draw[thick,dashed,->] (0.5,0) -- (0.5,-0.75) -- (6,-0.75);
     \draw[thick,dashed,->] (8.2,-0.75) -- (9,-0.75) node at (12,-0.5){$u^{(i)}[0,m], u^{(i)}[1,m], \cdots, u^{(i)}[K_i-1,m]$};
    \end{tikzpicture}
    }
    \caption{Model for a complex-valued c.c.s, that determines the signal $x^{(i)}[n,m]$ transmitted by the $i$-th TX/radar in Fig.~\ref{fig:fmcw_vs_ccs}a in the $m$-th slow time interval.}
    \label{fig:coding_setup}
\end{figure}
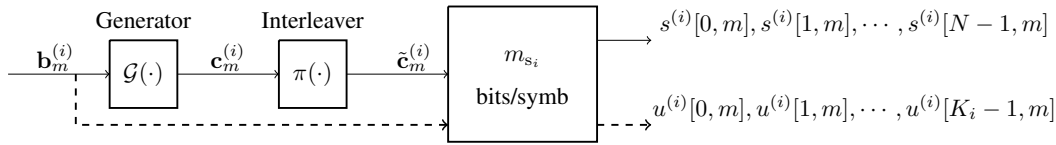
We model a c.c.s as shown in Fig.~\ref{fig:coding_setup}. At TX~$i$, for each $m$, a binary message vector, $\nbb_m^{(i)} \in \{0,1\}^{K_i m_{{\rm s}_i}}$, of length $K_i m_{{\rm s}_i}$ is transformed into a binary codeword vector, $\nbc_m^{(i)} \in \{0,1\}^{Nm_{{\rm s}_i}}$, of length $Nm_{{\rm s}_i}$ through a generator mapping, $\ncalG(\cdot)$\footnote{For linear block codes, $\ncalG(\cdot)$ takes the form of either a generator matrix or a parity check matrix.}. The encoded bits are then interleaved and mapped to a discrete, bounded constellation through a mapping that associates $m_{{\rm s}_i}$ bits per symbol to produce a complex-valued c.c.s block, $s^{(i)}[n,m]~(n=0,\cdots,N-1)$, of block length\footnote{In coding theory, the block length typically refers to the number of bits in the codeword vector, $\nbc_m^{(i)}$ (equal to $Nm_{{\rm s}_i}$). Since the quantities of interest in this paper take values in $\nbbC$ (e.g., the auto-correlation function), we slightly abuse the notation and refer to the number of complex-valued symbols in the signal $s^{(i)}[n,m]$ (for fixed $m$) as the block length.}, $N$. Let $u^{(i)}[n,m]~(n=0,\cdots,K_i-1)$ denote the complex-valued message signal, of length $K_i$, obtained by mapping $\nbb_m^{(i)}$ to the same constellation. The code rate, $r_i$, of $s^{(i)}[n,m]$ equals $K_i/N$. 

In general, each c.c.s symbol, $s^{(i)}[n,m]$, is a \emph{non-linear} function of one or more message symbols ($u^{(i)}[n_1,m]$, $u^{(i)}[n_2,m],\cdots$ etc.), which is difficult to characterize for most modulation and coding schemes\footnote{Even for linear codes, the relationship between the message and c.c.s symbols is non-linear, due to the symbol mapping operation.}. Thus, for tractability, we make some assumptions regarding the signals $u^{(i)}[n,m]$ and $s^{(i)}[n,m]$ below.
\begin{assumption}[i.i.d. zero-mean message symbols]
\label{assump:iid}
 The message symbols, $u^{(i)}[0,m], \cdots, u^{(i)}[K_i-1,m]$ are i.i.d with zero mean and finite energy, as they are typically drawn uniformly from bounded, symmetric constellations like QPSK, QAM, etc.
\end{assumption}

\begin{assumption}[Systematic Encoding]
\label{assump:sys_encoding}
 The first $K_i$ c.c.s symbols coincide with the message symbols, i.e., $s^{(i)}[n,m] = u^{(i)}[n,m]~(n=0,\cdots,K_i-1)$. This is referred to as systematic encoding and any linear block code can be transformed into systematic form through a linear mapping. Since widely used codes such as LDPC and Polar Codes are linear block codes, we believe the systematic encoding assumption\footnote{Note that we assume systematic encoding post-interleaving, as illustrated in Fig.~\ref{fig:coding_setup}. Hence, $\pi(k) = k$ for $k \in \{1,\cdots,Km_{{\rm s}_i}\}$, where $\pi(k)$ denotes the position of the $k$-th bit in $\nbc_m^{(i)}$ in the interleaved codeword, $\tilde{\nbc}_m^{(i)}$.} is reasonable. 
\end{assumption}

\begin{assumption}[Uncorrelated c.c.s symbols]
\label{assump:uncorr}
 While the c.c.s symbols, $s^{(i)}[0,m], \cdots, s^{(i)}[N-1,m]$, are statistically dependent, in general, we assume that they are mutually uncorrelated. While Assumption~\ref{assump:sys_encoding} ensures that the first $K_i$ symbols are uncorrelated, the encoding operation typically introduces correlation across subsequent codeword bits (i.e., in $\nbc_m^{(i)}$ in Fig.~\ref{fig:coding_setup}), which can be mitigated by interleaving; for instance, in Appendix~\ref{app:interleave}, we demonstrate that for repetition codes, the interleaved codeword bits ($\tilde{\nbc}_m^{(i)}$ in Fig.~\ref{fig:coding_setup}) are asymptotically uncorrelated for large $N$ (i.e., the correlation coefficient between a pair of arbitrarily chosen bits in $\tilde{\nbc}_m^{(i)}$ tends to 0, as $N \rightarrow \infty$). Hence, we believe this to be a reasonable assumption concerning c.c.s symbols for most practical codes.
 % \footnote{In a sense, the repetition code induces the ``strongest" correlation across certain codeword bits/symbols.}
 % \footnote{This result holds despite the fact that systematic encoding imposes some restrictions on the interleaver, $\pi(\cdot)$.}
\end{assumption}

The relationship between the signals $s^{(i)}[n,m]$ and $x^{(i)}[n,m]$ depends on whether the latter is a single- or multi-carrier waveform. We characterize this relationship for a representative of each type, namely the conventional single-carrier and the OFDM waveforms.

\subsubsection{Conventional Single-Carrier}
\label{subsec:ccs_sc}
The c.c.s symbols are transmitted as is in time-domain, i.e.,
\begin{IEEEeqnarray}{rCl}
\label{eq:sc_xs}
    x^{(i)}[n,m] & = & s^{(i)}[n,m].
\end{IEEEeqnarray}

\subsubsection{OFDM}
\label{subsec:ccs_ofdm}
The c.c.s symbols in $s^{(i)}[k,m]~(k=0,\cdots,N-1)$ are multiplexed in the frequency domain over $N$ subcarriers to constitute the $m$-th OFDM symbol. Thus, $x^{(i)}[n,m]$ and $s^{(i)}[k,m]$ are related by an IDFT, as follows:
\begin{IEEEeqnarray}{rCl}
\label{eq:ofdm_xs}
    x^{(i)}[n,m] & = & \frac{1}{N} \sum_{k=0}^{N-1} s^{(i)}[k,m] \exp \left(j 2 \pi \frac{n}{N} k \right). 
\end{IEEEeqnarray}
The total number of OFDM symbols transmitted equals $M$, as per Fig.~\ref{fig:fmcw_vs_ccs}.

\subsection{Sensing Model}
\label{subsec:sensing_model}
For the target scene in Fig.~\ref{fig:setup}a, let $(n_t^{(i)},m_t^{(i)}) \in \{1,\cdots,n_{\rm max}\} \times \{1,\cdots,M\}$ denote the (range, Doppler)-bin in which the $t$-th target $(t=1,\cdots,N_{\rm tar})$ is situated w.r.t the $i$-th TX/radar (i.e., corresponding to the path TX $i \rightarrow$ Target $\rightarrow$ TX $i$), where $n_{\rm max} \ll N$ denotes the range bin associated with the maximum detectable range. Then, the (discrete) time-domain radar return, $y^{(i)}[n,m]$, at the $i$-th TX/radar corresponding to the transmission of $x^{(i)}[n,m]$ is given by:
\begin{IEEEeqnarray}{rCl}
\label{eq:basic_radar_return}
    y^{(i)}[n,m] & = & \sum_{t=1}^{N_{\rm tar}} \underbrace{\alpha_t^{(i)} x^{(i)}[n-n_t^{(i)},m] \exp\left(j 2\pi \frac{m_t^{(i)}}{M} (m-1) \right)}_\text{Desired radar returns} + \underbrace{y_{\rm intf}^{(i)}[n,m]}_\text{Sensing intf.} + w^{(i)}[n,m],
\end{IEEEeqnarray}
where 
\begin{itemize}
    \item $\alpha_t^{(i)} \in \nbbC$ is the gain, capturing the combined effects of beamforming, path loss, and target reflection/scattering along the path TX $i \rightarrow$ Target $t \rightarrow$ TX $i$;
    
    \item  $y_{\rm intf}^{(i)}[n,m]$ denotes the \emph{sensing interference} signal at the $i$-th TX/radar; and,
    
    \item $w^{(i)}[n,m]\sim \ncalC\ncalN(0,\sigma^2)$ is the additive noise signal, independent of the radar returns and also across $n,m$.
\end{itemize}
In the terms capturing the desired radar returns in (\ref{eq:basic_radar_return}), the delay shift corresponding to the target ranges is assumed to be the same across all $m$, and the Doppler shift is assumed to be constant over the duration of a slow-time interval. Such a decoupling between the delay and Doppler shifts can be assumed when the Doppler frequency is much smaller than the signal bandwidth.%\footnote{For a signal bandwidth $B = 1{\rm GHz}$ (enabling a range resolution of $15{\rm cm}$) and block length, $N = 1000$, the slow-time duration is $N/B = 1{\rm \mu s}$ . At a carrier frequency of $100{\rm GHz}$, a target moving at $30{\rm mph}$ yields a maximum Doppler frequency of $\approx 9{\rm kHz}$. The variation in the phase of a $9{\rm kHz}$ sinusoid over $1{\rm \mu s}$ is negligible.}

We model $y^{(i)}_{\rm intf}[n,m]$ as follows: let $y_{\rm intf}^{(q \rightarrow i)}[n,m]$ denote the sensing interference experienced at the $i$-th TX/radar due to the transmissions of TX/radar $q~(\neq i)$, which can be expressed as a superposition of $N_{\rm tar}+1$ components as follows:
\begin{IEEEeqnarray}{rCl}
\label{eq:y_intf_pair}
    y_{\rm intf}^{(q \rightarrow i)}[n,m] & = & \sum_{l=0}^{N_{\rm tar}} \alpha_l^{(q \rightarrow i)} x^{(q)}[n-n_l^{(q \rightarrow i)},m] \exp\left(j 2\pi \frac{m_l^{(q \rightarrow i)}}{M} (m-1) \right),
\end{IEEEeqnarray}
where the $l$-th term in (\ref{eq:y_intf_pair}) represents the interfering signal along the path TX $q \rightarrow $ Target $l \rightarrow$ TX $i$ for $l\neq 0$, while the term corresponding to $l=0$ represents the signal leakage along the direct path TX $q \rightarrow $ TX $i$. Similar to the desired radar returns in (\ref{eq:basic_radar_return}), each such path is parameterized by a gain, delay and Doppler shift, captured by $\alpha_l^{(q \rightarrow i)} \in \nbbC$, $n_l^{(q \rightarrow i)} \in \{1,\cdots,n_{\rm max}\}$ and $m_l^{(q \rightarrow i)}\in \{1,\cdots,M\}$, respectively. Thus, $y^{(i)}_{\rm intf}[n,m]$ can be expressed as follows:
\begin{IEEEeqnarray}{rCl}
\label{eq:y_intf_total}
    y^{(i)}_{\rm intf}[n,m] & = & \sum_{q \neq i} y^{(q \rightarrow i)}_{\rm intf}[n,m]. 
\end{IEEEeqnarray}
We now explore the sensing signal processing for the conventional single-carrier and the OFDM waveforms.

\subsubsection{Conventional Single-Carrier}
\label{subsubsec:sensing_sc}
For a single-carrier waveform, the target range bins can be estimated from (\ref{eq:basic_radar_return}) using a bank of matched filters, followed by a DFT to estimate the Doppler bins, as described below. Let $r^{(i)}[l,m]~(l = 1,\cdots, n_{\rm max})$ denote the (normalized) matched filter output, corresponding to $y^{(i)}[n,m]$ and $x^{(i)}[n,m]$, given by:
\begin{IEEEeqnarray}{rCl}
\label{eq:mf_bank}
    r^{(i)}[l,m] & := & \frac{1}{N}\sum_{n=-\infty}^{\infty} y^{(i)}[n,m] x^{(i)^*}[n-l,m] \notag \\ %, ~(l=0,\cdots,n_{\rm max} \ll N-1) 
               & = &\underbrace{\sum_{t=1}^{N_{\rm tar}} \alpha_t^{(i)} \chi(l-n_t^{(i)};s^{(i)}[\cdot,m]) \exp\left(j 2\pi \frac{m_t^{(i)}}{M}(m-1)\right)}_\text{Desired signal component} \notag \\
               & & + \underbrace{\sum_{q \neq i} \sum_{ t' = 0}^{N_{\rm tar}} \alpha_{t'}^{(q \rightarrow i)} \varrho(l-n_{t'}^{(q \rightarrow i)};s^{(q)}[\cdot,m],s^{(i)}[\cdot,m]) \exp \left( j 2\pi \frac{m_{t'}^{(q \rightarrow i)}}{M}(m-1) \right)}_\text{Sensing interference component} \nonumber \\
               & & + \frac{1}{N} \sum_{n=-\infty}^{\infty} w^{(i)}[n,m] s^{(i)^*}[n-l,m].% \\
%\label{eq:mf_sc_noise}               
%        \mbox{where}~ \tilde{w}^{(i)}[l,m] &:=  
\end{IEEEeqnarray}
Eqn. (\ref{eq:mf_bank}) is obtained from (\ref{eq:sc_xs}) and Definitions~\ref{def:autocorr} and \ref{def:crosscorr}.

\begin{nrem}[Sensing Interference Management for a single-carrier waveform]
\label{rem:intf_supp_sc}
 The sensing interference in (\ref{eq:mf_bank}) can be completely eliminated if and only if the collection of c.c.s transmitted by the TXs have mutually zero cross-correlation across all lags [i.e., $\varrho(l; s^{(q)}[\cdot,m],s^{(i)}[\cdot,m]) \equiv 0,~ \forall q\neq i$], which is unlikely for all realizations of random c.c.s, $s^{(i)}[n,m]$ and $s^{(q)}[n,m]$. In particular, a large value of $|\varrho(l;s^{(i)}[\cdot,m],s^{(q)}[\cdot,m])|$ at a non-target range bin, $l$, gives rise to undesired \underline{sidelobes} in $|r^{(i)}[l,m]|$ which, in turn, affects the sensing performance by giving rise to false-alarms and missed detections -- the latter due to the \underline{near-far effect}, wherein the weaker peaks of farther targets are buried among the stronger interference sidelobes. To minimize the occurrence of these outcomes, it is desirable for $\nbbP(|\varrho(l; s^{(i)}[\cdot,m], s^{(q)}[\cdot,m])|> u)$ -- the tail probability of $|\varrho(l; s^{(i)}[\cdot,m], s^{(q)}[\cdot,m])|$ -- to decay rapidly w.r.t $u$. We characterize this quantity in terms of the c.c.s parameters (i.e., $r_i$, $r_q$ and $N$) in Section~\ref{subsec:tail_prob_cc}.
\end{nrem}

The single-carrier range-Doppler map, $R^{(i)}_{\rm sc}[l,\nu]$, whose magnitude captures the strength of the radar return from the $l$-th range bin and $\nu$-th Doppler bin, can be obtained from $r^{(i)}[l,m]$ through a $M$-point DFT across the slow time index, $m$. Assuming the condition in Remark~\ref{rem:intf_supp_sc} holds, the resulting expression for $R^{(i)}_{\rm sc}[l,\nu]$ is as follows:
\begin{align}
\label{eq:rd_map}
   R_{\rm sc}^{(i)}[l,\nu] &:= \sum_{m=1}^M r^{(i)}[l,m] \exp\left(-j 2\pi \frac{\nu}{M} (m-1) \right) \notag \\
   &= \sum_{t=1}^{N_{\rm tar}} \alpha_t^{(i)}  \sum_{m=1}^M |\chi(l-n_t^{(i)};s^{(i)}[\cdot,m])| \exp \left[j \left( 2 \pi \left(\frac{m_t^{(i)}-\nu}{M} \right)(m-1) + \underbrace{\arg\left(\chi(l-n_t^{(i)};s^{(i)}[\cdot,m])\right)}_\text{autocorr.~phase noise} \right) \right]
\end{align}
   
\begin{nrem}[Ideal Auto-correlation]
\label{rem:ideal_autocorr}
It is easy to see from (\ref{eq:rd_map}) that at a large enough SNR, if $\chi(l;s^{(i)}[\cdot,m]) \propto \delta[l]~ \forall m$ (i.e., ideal auto-correlation function), then $R_{\rm sc}^{(i)}[l,\nu] \propto \sum_{t=1}^{N_{\rm tar}} \alpha_t^{(i)} \delta[l-n_t^{(i)}] \delta[\nu-m_t^{(i)}]$, resulting in sharp peaks for $|R_{\rm sc}^{(i)}[l,\nu]|$ at $(l,\nu) \in \{(n_t^{(i)},m_t^{(i)}): t=1,\cdots,N_{\rm tar} \}$ -- the targets' range-Doppler bins w.r.t TX~$i$. However, an ideal auto-correlation function is unlikely for all realizations of random c.c.s, $s^{(i)}[n,m]$, and when $\chi(l;s^{(i)}[\cdot,m]) \not\propto \delta[l]$, 
\begin{itemize}
    \item[(a)] a large value of $|\chi(l;s^{(i)}[\cdot,m])|$ at $l \neq 0$ (i.e., sidelobe) can give rise to false-alarms and the near-far effect, similar to Remark~\ref{rem:intf_supp_sc}; and,
    \item[(b)] the phase of $\chi(l;s^{(i)}[\cdot,m])$ impacts Doppler-bin estimation, as seen in (\ref{eq:rd_map}).
\end{itemize}
To minimize the occurrence of these outcomes, it is desirable for $\nbbP(|\chi(l; s^{(i)}[\cdot,m])|> u)$ -- the tail probability of $|\chi(l; s^{(i)}[\cdot,m])|~(l\neq 0)$ -- to decay rapidly w.r.t $u$. We characterize this quantity in terms of the c.c.s parameters (i.e., $r_i$ and $N$) in Section~\ref{subsec:tail_prob_ac}.  
\end{nrem}

%\begin{nrem}[Sensing using c.c.s with a single-carrier waveform]
%\label{rem:sensing_ccs_sc}
%From Remarks~\ref{rem:intf_supp_sc} and \ref{rem:ideal_autocorr}, we observe that for single-carrier waveforms, the collection of modulating c.c.s, $\{s^{(i)}[\cdot,m]: i=1,\cdots,M_r\}$, must have ideal correlation properties to simultaneously achieve high sensing, while suppressing mutual sensing interference. 
%\end{nrem}
% wherein, the use of distinct $s_m[\cdot]$ leads to improved spectral efficiency, at the expense of a slight increase in signal processing complexity.

\subsubsection{OFDM}
\label{subsubsec:sensing_ofdm}
% \footnote{Although matched filtering can also be used for the OFDM waveform, the frequency domain approach of (\ref{eq:ofdm_range_sinusoid}) is widely adopted, as it yields a higher peak-to-sidelobe ratio \cite{Sturm_Wiesbeck_2011}.}
The OFDM radar processing chain at the $i$-th TX involves an $N$-point DFT of $y^{(i)}[n,m]$ in (\ref{eq:basic_radar_return}) over the fast time index, $n$, to obtain a frequency-domain signal, $Y^{(i)}[k,m]~(k=0,\cdots,N-1)$, that has the following expression:
\begin{IEEEeqnarray}{rCl}
\label{eq:ofdm_range_sinusoid}
    Y^{(i)}[k,m] & = & \underbrace{\sum_{t=1}^{N_{\rm tar}} \alpha_t^{(i)} s^{(i)}[k,m] \exp\left(-j 2\pi \frac{n_t^{(i)}}{N} k \right) \exp\left(j 2\pi \frac{m_t^{(i)}}{M} (m-1) \right)}_\text{Desired signal returns}  \notag \\
    & & + \underbrace{\sum_{q \neq i} \sum_{l=0}^{N_{\rm tar}} \alpha_l^{(q \rightarrow i)} s^{(q)}[k,m] \exp \left(-j 2\pi \frac{n_l^{(q \rightarrow i)}}{N} k \right) \exp \left(j 2\pi \frac{m_l^{(q \rightarrow i)}}{M} (m-1) \right)}_\text{Interfering signal returns} \nonumber \\
    & & + \sum_{n=0}^{N - 1} w^{(i)}[n,m] \exp\left(-j 2\pi \frac{k}{N} n \right).
\end{IEEEeqnarray}
At the $i$-th TX, upon dividing (\ref{eq:ofdm_range_sinusoid}) by the known $s^{(i)}[k,m]$, we obtain\footnote{The LHS of (\ref{eq:ofdm_radar_setup}) corresponds to zero-forcing equalization, which we assume in our OFDM analysis in this paper. With minor modifications, our analysis holds for other forms of equalization, as well (e.g., MMSE).}:
\begin{IEEEeqnarray}{rCl}
\label{eq:ofdm_radar_setup}
    \frac{Y^{(i)}[k,m]}{s^{(i)}[k,m]} & = & \underbrace{\sum_{t=1}^{N_{\rm tar}} \alpha_t^{(i)} \exp\left(-j 2\pi \frac{n_t^{(i)}}{N} k \right) \exp\left(j 2\pi \frac{m_t^{(i)}}{M} (m-1) \right)}_\text{Desired signal component}  \notag \\
    & & + \underbrace{\sum_{q \neq i} \sum_{l=0}^{N_{\rm tar}} \alpha_l^{(q \rightarrow i)} \frac{s^{(q)}[k,m]}{s^{(i)}[k,m]} \exp \left(-j 2\pi \frac{n_l^{(q \rightarrow i)}}{N} k \right) \exp \left(j 2\pi \frac{m_l^{(q \rightarrow i)}}{M} (m-1) \right)}_\text{Sensing interference component} \nonumber \\
    & & + \frac{W^{(i)}[k,m]}{s^{(i)}[k,m]}.
\end{IEEEeqnarray}
%\footnote{The OFDM cyclic prefix causes a scaling of the (discrete) Doppler frequencies in (\ref{eq:ofdm_radar_setup}), but we ignore this for convenience.}
For the desired signal component in (\ref{eq:ofdm_radar_setup}), each of the terms under the summation sign is a scaled product of a pair of decoupled sinusoids -- one across the subcarriers (index $k$), whose frequency depends on the target ranges, and another across OFDM symbols (index $m$), whose frequency depends on the target velocity (Doppler). Hence, the target parameters can be estimated from the OFDM range-Doppler map, $R^{(i)}_{\rm OFDM}[l,\nu]$, obtained through a combination of an $N$-point IDFT (over $k$) and an $M$-point DFT (over $m$), as follows  \cite{Sturm_Wiesbeck_2011}:
\begin{align}
    \label{eq:ofdm_rd_map}
    R^{(i)}_{\rm OFDM}[l,\nu] &:= \frac{1}{N} \sum_{m=1}^M \sum_{k=0}^{N-1} \frac{Y^{(i)}[k,m]}{s^{(i)}[k,m]} \exp\left(j 2\pi \frac{l}{N}k\right) \exp\left(-j 2\pi \frac{\nu}{M}(m-1) \right) \notag \\
    &= \underbrace{\sum_{t=1}^{N_{\rm tar}} \alpha_t^{(i)}\delta\left[l-n_t^{(i)}\right]\delta\left[\nu - m_t^{(i)}\right]}_\text{Peaks at $(l,\nu) = \{(n_t^{(i)},m_t^{(i)}): t=1,\cdots,N_{\rm tar}\}$} \notag \\
    &~+ \underbrace{\sum_{q \neq i} \sum_{l=0}^{N_{\rm tar}} \alpha_l^{(q \rightarrow i)} \sum_{m=1}^M \exp \left(j 2\pi \frac{(m_l^{(q \rightarrow i)}-\nu)}{M} (m-1) \right). \underbrace{\frac{1}{N} \sum_{k=0}^{N-1}  \frac{s^{(q)}[k,m]}{s^{(i)}[k,m]} \exp \left(j 2\pi \frac{(l - n_l^{(q \rightarrow i)})}{N} k \right)}_\text{IDFT of $s^{(q)}[k,m]/s^{(i)}[k,m]$}}_\text{Distortion induced by sensing intf.}  \notag \\
    &~ + \underbrace{\tilde{W}^{(i)}[l,\nu]}_\text{Noise component}.
\end{align}
From (\ref{eq:ofdm_rd_map}), it is clear that in the absence of interference and a large enough SNR, $|R^{(i)}_{\rm OFDM}[l,\nu]|$ would have sharp peaks only at $(l,\nu) \in \{(n_t^{(i)},m_t^{(i)}): t = 1,\cdots,N_{\rm tar}\}$ -- the targets' range-Doppler bins w.r.t TX~$i$. Furthermore, we also observe that the distortion in $R^{(i)}_{\rm OFDM}[l,\nu]$ due to the interference from $s^{(q)}[k,m]$ is governed by the $N$-point IDFT of $s^{(q)}[k,m]/s^{(i)}[k,m]$ over the index $k$. We remark on this quantity below.

\begin{ndef}
Let
\begin{align}
\label{eq:def_V}
    V^{(q \rightarrow i)}[l,m] &:= \frac{1}{N} \sum_{k=0}^{N-1} \frac{s^{(q)}[k,m]}{s^{(i)}[k,m]} \exp\left(j 2\pi \frac{l}{N} k \right)
\end{align}
denote the $N$-point IDFT of $s^{(q)}[k,m]/s^{(i)}[k,m]~(q\neq i)$ over the index $k$.
\end{ndef}

\begin{nrem}[Distortion due to sensing interference in the OFDM range-Doppler map]
We may reasonably assume that for fixed $m$, the signals $s^{(i)}[k,m]$ and $s^{(q)}[k,m]$ are independent, as the c.c.s from different TXs are generated from mutually independent message signals. Furthermore, since $s^{(i)}[k,m] \neq 0$, it follows from Assumption~\ref{assump:iid} that $\nbbE\left[\frac{s^{(q)}[k,m]}{s^{(i)}[k,m]}\right] = 0$. Hence, $V^{(q \rightarrow i)}[l,m]$ also has zero mean, and the interference term in (\ref{eq:ofdm_rd_map}) can be viewed as a zero-mean, additive \underline{non-Gaussian} distortion.
\end{nrem}

\begin{nrem}[Sensing interference management for an OFDM waveform]
\label{rem:intf_supp_ofdm}
Despite being zero-mean, the distortion due to the sensing interference in (\ref{eq:ofdm_rd_map}) can give rise to sidelobes in $|R^{(i)}_{\rm OFDM}[l,\nu]|$, accompanied by false-alarms and the near-far effect (similar to Remark~\ref{rem:intf_supp_sc}), if $|V^{(q \rightarrow i)}[l,m]|$ takes on a large value at a non-target range bin, $l$. To minimize the occurrence of these outcomes, it is desirable for $\nbbP(|V^{(q \rightarrow i)}[l,m]| > u)$ -- the tail probability of $|V^{(q \rightarrow i)}[l,m]|$ -- to decay rapidly w.r.t $u$. We characterize this quantity in terms of the c.c.s parameters (i.e., $r_i$, $r_q$ and $N$) in Section~\ref{subsec:tail_prob_ofdm}.
\end{nrem}

From Remarks~\ref{rem:intf_supp_sc} through \ref{rem:intf_supp_ofdm}, it is clear that the sensing potential of c.c.s is intricately linked to the tail probabilities of $|\varrho(l; s^{(i)}[\cdot,m], s^{(q)}[\cdot,m])|$, $|V^{(q \rightarrow i)}[l,m]|$ and $|\chi(l;s^{(i)}[\cdot,m])|~(l\neq 0)$ -- the first two quantities determine the sensing interference suppression capabilities of c.c.s in single-carrier and OFDM waveforms, respectively, while the last impacts the sensing performance of c.c.s in a single-carrier waveform. We characterize these tail probabilities in the following section.

\section{Sensing Potential of c.c.s}
\label{sec:corr}
Consider a pair of c.c.s blocks, $s^{(i)}[n,m]$ and $s^{(q)}[n,m]~(i\neq q)$, generated by message signals, $u^{(i)}[n,m]$ and $u^{(q)}[n,m]$, respectively, according to Section~\ref{subsec:ccs_model}. Since $m$ is arbitrary, we omit this index throughout this section to simplify the notation for all the quantities of interest, such as $\chi(l;s^{(i)}[\cdot])$, $\varrho(l, s^{(i)}[\cdot],s^{(q)}[\cdot])$, $V^{(q \rightarrow i)}[l]$,  etc. In deriving bounds for the tail probabilities of the latter, we restrict our attention to their real parts; the analysis for the imaginary part follows similarly. We now introduce a few lemmas that will be used in deriving our main results later in this section.

\begin{lem}[Hoeffding's Lemma]
\label{lem:hoeff}
 Let $X$ be a (real) random variable, satisfying $|X|\leq b < \infty$ with probability one, and $\nbbE[X]=0$. Then, for $\lambda > 0$, 
 \begin{IEEEeqnarray}{rCl}
     \nbbE\left[\exp(\lambda X)\right] & \leq & \exp\left(\lambda^2 b^2/2\right).
 \end{IEEEeqnarray}
\end{lem}
\begin{IEEEproof}
 See \cite[Lemma~2.19]{Bercu_Deylon_Rio_2015}. %Appendix~\ref{app:lem_Hoeffding}.
\end{IEEEproof}

\begin{comment}
\begin{lem}[Hoeffding's Inequality]
\label{lem:hoeff_ineq}
 For a collection of independent (real) random variables, $X_1,\cdots,X_n$, satisfying $\nbbE[X_i] = 0$ and $|X_i| \leq b < \infty$ with probability one for all $i$, and real weights $w_1,\cdots,w_N$,
 \begin{IEEEeqnarray}{rCl}
     \nbbP\left(\left|\sum_{i=1}^N w_i X_i \right| > u\right) & \leq & 2 \exp\left(- \frac{Nu^2}{2b^2 \sum_{i=1}^N w_i^2} \right)
 \end{IEEEeqnarray}
\end{lem}
\begin{IEEEproof}
 See \cite[Theorem~2.16]{Bercu_Deylon_Rio_2015}.
\end{IEEEproof}
\end{comment}

\begin{lem}
\label{lem:am-gm}
For $\lambda \in \nbbR$, and a collection of (real) identically distributed (but not necessarily independent) random variables, $X_1, \cdots, X_n$
 \begin{IEEEeqnarray}{rCl}
 \label{eq:am-gm_ineq}
\nbbE\left[\exp\left(\lambda X_1 + \cdots + \lambda X_n \right)\right] & \leq & \nbbE\left[\exp \left(\lambda n X_1\right)\right].
 \end{IEEEeqnarray}
\end{lem} 
\begin{IEEEproof}
From the AM-GM inequality, we have for $\lambda \in \nbbR$,
 \begin{IEEEeqnarray}{rCl}
  \exp \left(\lambda X_1 + \cdots + \lambda X_n \right) & \leq & \frac{\exp \left(\lambda nX_1\right) + \cdots + \exp \left(\lambda nX_n\right)}{n}
 \end{IEEEeqnarray}
Applying the $\nbbE[\cdot]$ operator to both sides of the above inequality, we obtain (\ref{eq:am-gm_ineq}).
\end{IEEEproof}

\subsection{Tail Probability of $|\Re \{\chi(l;s[\cdot])\}|$}
\label{subsec:tail_prob_ac}
To further simplify the notation, we omit the index $i$ in this subsection, and denote respectively by $\chi(l;s[\cdot])$ and $u[n]$, the auto-correlation function and the message signal associated with c.c.s, $s[n]$.
\begin{ndef}
\label{def:Ys}
Let $M_l:=\left\lceil \frac{N-l}{K-l} \right\rceil$. Then, $\Re\{\chi(l;s[\cdot])\}$ can be expressed as follows :
\begin{IEEEeqnarray}{rCl}
    \Re\{\chi(l;s[\cdot])\} &:= & \frac{1}{N} \left(Y_1^{(l)} + \cdots + Y_{M_l}^{(l)}\right), \notag\\
\label{eq:Y_first_l}    
\mbox{where} ~ Y_i^{(l)} & := & \Re\left\{ \sum_{k'=0}^{K-1-l} s[(i-1)(K-l)+k'+l] s^*[(i-1)(K-l)+k'] \right\} \nonumber \\
& & (i=1,\cdots,M_l-1), \\
\label{eq:Y_last_l}
\mbox{and} ~ Y_{M_l}^{(l)} & := & \Re\left\{ s\left[\left(M_l -1\right)(K-l)+l\right] s^*\left[ \left(M_l -1 \right)(K-l) \right] \right.  \notag \\
 & & \left. + s\left[\left(M_l -1\right)(K-l)+l+1\right] s^*\left[ \left(M_l -1 \right)(K-l)+1 \right] \right.  \notag \\ 
 & & \left. + \cdots + s[N-1] s^*[N-1-l] \right\}. 
\end{IEEEeqnarray} 
According to (\ref{eq:Y_first_l}), $Y_1^{(l)} := \Re \left\{s[l]s^*[0] + s[l+1] s^*[1] + \cdots + s[K-1]s^*[K-1-l] \right\}$ denotes the sum of the first $(K-l)$ non-trivial terms in the RHS of (\ref{eq:chi_s}), $Y_2^{(l)}$ denotes the sum of the next $(K-l)$ non-trivial terms, and so on. The last such quantity, $Y_{M_l}^{(l)}$, may contain fewer terms, but for simplicity, we assume that it contains $K-l$ non-trivial terms, as well. Decomposing $\Re\{\chi(l;s[\cdot])\}$ into partial sums in this manner, along with Assumption~\ref{assump:sys_encoding}, ensures that $Y_1^{(l)}$ is a function of i.i.d random variables (i.e., the message signal $u[n]$), whose distribution can be characterized using concentration inequalities, as shown in Lemma~\ref{lem:charY1} below. On the other hand, $Y_i^{(l)}~(i\neq 1)$, is much harder to characterize, as it is a function of a mixture of i.i.d and dependent random variables, in general. Thus, to help characterize $Y_i^{(l)}~(i\neq 1)$, we make the following assumption.
\end{ndef}

\begin{assumption}
\label{assump:id}
Due to their similarity in form, we assume that $Y_1^{(l)}, \cdots, Y_{M_l}^{(l)}$ are identically distributed. This is certainly true for $(N,K)$-linear MDS (Maximum Distance Separable) codes (e.g., the Reed-Solomon code) since any collection of $K$ c.c.s symbols are mutually i.i.d. for such codes \cite[Theorem~5.4.5]{coding_theory_book}, and each $Y_i^{(l)}$ is a function of $K$ c.c.s symbols, namely, $s[(i-1)(K-l)], s[(i-1)(K-l)+1], \cdots, s[i(K-l)+l-1)]$. Note that $Y_1^{(l)}, \cdots, Y_{M_l}^{(l)}$ are statistically dependent, in general.
\end{assumption}

\begin{lem}
\label{lem:charY1}
For $\lambda \in \nbbR$,
\begin{IEEEeqnarray}{rCl}
    \nbbE\left[\exp\left( \lambda Y_1^{(l)} \right)\right] & \leq & \exp\left(\lambda^2 b^2(K-l)/2 \right),
\end{IEEEeqnarray}
where $b$ satisfies $|\Re \{s[n] s^*[n-l]\}| \leq b < \infty$, for any $n,l$.
\end{lem}
\begin{IEEEproof}
Let $\nbu^{(K-2)} := \left[u[0], \cdots, u[K-2]\right]$. From Assumption~\ref{assump:sys_encoding} and (\ref{eq:Y_first_l}), we have
 \begin{IEEEeqnarray}{rCl}
     Y_1^{(l)} & = & \Re\left\{u[l] u^*[0] + u[l+1] u^*[1] + \cdots + u[K-1]u^*[K-1-l] \right\} \notag \\
     \label{eq:A_motivation}
     & = & A(\nbu^{(K-2}) + B(u[K-1]), \\
     \label{eq:Adef}
     \mbox{where}~ A(\nbu^{(K-2}) & := & \Re\left\{u[l] u^*[0] + u[l+1]u^*[1] + \cdots + u[K-2]u^*[K-2-l] \right\}, \\
     \label{eq:Bdef}
     \mbox{and}~ B(u[K-1]) &:= &\Re\left\{u[K-1] u^*[K-1-l] \right\}.
 \end{IEEEeqnarray}
Conditioning on $\nbu^{(K-2)}$, $\nbbE\left[\exp\left( \lambda Y_1^{(l)} \right) \right]$ can be written as:
 \begin{align}
    \label{eq:repE}
    \nbbE\left[\exp \left( \lambda Y_1^{(l)} \right)\right] &= \nbbE_{\nbu^{(K-2)}}\left[ \nbbE_{u[K-1]}\left[\exp\left( \lambda Y_1^{(l)} \right) | \nbu^{(K-2)}\right] \right] \notag \\
    &= \nbbE_{\nbu^{(K-2)}}\left[ \exp \left( \lambda A(\nbu^{(K-2)}) \right) \nbbE_{u[K-1]} \left[\exp \left(\lambda B(u[K-1]) \right) | \nbu^{(K-2)}\right] \right]. 
 \end{align}
Since $u[K-1]$ is independent of $\nbu^{(K-2)}$, we have
\begin{align}
\label{eq:hoeff_apply}
    \nbbE_{u[K-1]} \left[\exp \left(\lambda B(u[K-1]) \right) | \nbu^{(K-2)}\right] = \nbbE_{u[K-1]} \left[\exp \left(\lambda B(u[K-1]) \right)\right]\leq \exp \left(\frac{\lambda^2 b^2}{2} \right),
\end{align}
where the latter inequality stems from Lemma~\ref{lem:hoeff}, as $|\Re\{u[K-1]u^*[K-1-l]\}| = |\Re\{s[K-1]s^*[K-1-l]\}| \leq b $. Thus, from (\ref{eq:repE}) and (\ref{eq:hoeff_apply}),
\begin{align}
    \nbbE\left[\exp \left( \lambda Y_1^{(l)} \right)\right] \leq \exp \left(\frac{\lambda^2 b^2}{2} \right) \nbbE_{\nbu^{(K-2)}}\left[ \exp \left( \lambda A(\nbu^{(K-2)}) \right)\right].
\end{align}
Repeating the above steps recursively by conditioning on $\nbu^{(K-3)},\nbu^{(K-4)}$ and so on, we get,
\begin{align}
\label{eq:hoeff_apply_full}
    \nbbE\left[\exp \left(\lambda Y_1^{(l)} \right)\right] \leq \exp\left(\frac{\lambda^2 b^2}{2}(K-l) \right) 
\end{align}
\end{IEEEproof}

\begin{thm}[Auto-correlation Upper Bound]
\label{thm:main}
 For $u > 0$, the tail probability of $|\Re\{\chi(l;s[\cdot])\}|~(l \neq 0)$ satisfies the following upper bound:
 \begin{align}
\label{eq:thm_main_ineq}
 \nbbP \left( |\Re \{\chi(l;s[\cdot])\}| > u \right) \leq 2\exp(-O(rN) u^2),
 \end{align}
as $K, N \rightarrow \infty$ and $K/N=r$.
\end{thm}
\begin{IEEEproof}
 We use the Chernoff bound to prove our result. For $u > 0$,
\begin{align}
    \nbbP \left( \Re \{\chi(l;s[\cdot])\} > u \right) &= \nbbP \left( Y_1^{(l)} + \cdots + Y_{M_l}^{(l)} > Nu \right) \hspace{3mm} (\mbox{from Definition~\ref{def:Ys}}) \notag \\
    &= \nbbP\left(\exp \left[ \lambda Y_1^{(l)}+ \cdots + \lambda Y_{M_l}^{(l)} \right] > \exp(\lambda Nu)\right) \hspace{3mm} (\mbox{for}~\lambda>0) \notag \\
    \label{eq:markov}
    &\leq \exp(-\lambda Nu) ~\nbbE \left[\exp \left(\lambda Y_1^{(l)}+ \cdots + \lambda Y_{M_l}^{(l)} \right)\right] \hspace{3mm} (\mbox{Markov ineq.}) \\
    \label{eq:am-gm-apply}
    &\leq \exp(-\lambda Nu) ~\nbbE \left[\exp \left(\lambda M_l Y_1^{(l)} \right)\right] \hspace{3mm} (\mbox{Lemma~\ref{lem:am-gm}}) \\
    \label{eq:chernoff_lambda}
    &\leq \exp\left(-\lambda Nu + \frac{\lambda^2 M_l^2 b^2}{2}(K-l) \right)\hspace{3mm} (\mbox{Lemma~\ref{lem:charY1}}).
\end{align}
The tightest upper bound in (\ref{eq:chernoff_lambda}) is obtained at $\lambda = Nu/(b^2 M_l^2 (K-l))$, by equating the derivative of the argument of the exponent to zero. Substituting this value in (\ref{eq:chernoff_lambda}), we obtain
\begin{align}
\label{eq:+veu}
    \nbbP \left( \Re \{\chi(l;s[\cdot])\} > u \right) \leq \exp\left(- \frac{N^2 u^2}{2b^2 M_l^2 (K-l) } \right) .
\end{align}
Similarly, 
\begin{align}
    \nbbP \left( \Re \{\chi(l;s[\cdot])\} < -u \right) &= \nbbP(Y_1^{(l)} + \cdots + Y_{M_l}^{(l)} < -Nu) \notag \\
    &= \nbbP\left(\exp(\lambda Y_1^{(l)} + \cdots + Y_{M_l}^{(l)}) > \exp(-\lambda Nu)\right) \hspace{5mm} (\mbox{for}~\lambda<0).
\end{align}
The rest of the analysis follows along the same lines as (\ref{eq:markov})-(\ref{eq:chernoff_lambda}), with $\lambda = -Nu/(b^2M_l^2(N-l))$, resulting in the following bound
\begin{align}
\label{eq:-veu}
    \nbbP \left( \Re \{\chi(l;s[\cdot])\} < -u \right) &\leq \exp\left(- \frac{N^2 u^2}{2b^2 M_l^2 (K-l) } \right) .
\end{align}
Combining (\ref{eq:+veu}) and (\ref{eq:-veu}), we get
\begin{align}
    \nbbP(|\Re\{\chi(l;s[\cdot])\}| > u) \leq 2\exp \left(- \frac{N^2 u^2}{2b^2 M_l^2 (K-l)} \right).
\end{align}
As $K,N \rightarrow \infty$ with $K/N = r$, $N^2/(2b^2 M_l^2 (K-l)) = O(rN)$. This completes the proof.
\end{IEEEproof}

\begin{nrem}[``Extent of i.i.d-ness"]
\label{rem:iid_extent}
To provide intuition for the result in Theorem~\ref{thm:main}, consider $r=1$, corresponding to an uncoded communications signal comprising i.i.d zero-mean symbols. For this special case, (\ref{eq:thm_main_ineq}) can be interpreted as a concentration inequality closely related to the Hoeffding inequality \cite[Theorem~2.16]{Bercu_Deylon_Rio_2015}, for which the order of the exponential decay is determined by the number of i.i.d random variables present (i.e., $N$). A c.c.s can be viewed as a mixture of i.i.d and dependent random variables/symbols; in particular, the maximal number of i.i.d symbols over a block length $N$ equals\footnote{This is true for any linear code, and does not require systematic encoding (i.e., Assumption~\ref{assump:sys_encoding}).} $K=rN$, which, in turn, governs the $O(rN)$ exponential decay in (\ref{eq:thm_main_ineq}).
\end{nrem}

\begin{thm}[Auto-correlation Lower Bound]
\label{thm:lb}
For linear codes, the tail probability of $|\Re\{\chi(l;s[\cdot])\}|~(l \neq 0,~l \ll N)$ satisfies the following lower bound for \underline{sufficiently small} $u > 0$:
 \begin{align}
\nbbP \left( |\Re (\chi(l;s[\cdot]))| > u \right) \geq \exp(-O(rN)), 
 \end{align}
where $K/N=r$. 
\end{thm}
\begin{IEEEproof}
Let $\ncalS$ denote the set of feasible values of the signal $s[n]$ (i.e., the codebook). In particular, let $a_0[n]$ denote the signal corresponding to the all-zero codeword (i.e., $\nbc = \mathbf{0} = \tilde{\nbc}$ in Fig.~\ref{fig:coding_setup}), which occurs with probability $2^{-m_s K}$ for linear codes. It is easily seen from (\ref{eq:chi_s}) that when all the symbols in $s[n]$ are identical, as is the case with $a_0[n]$, then $|\chi(l;s[\cdot])| > u$ for $l \ll N$ and sufficiently small $u$. Thus, conditioning on $\ncalS$, we have
 \begin{align}
     \nbbP \left( |\Re (\chi(l;s[\cdot]))| > u \right) &= \sum_{a[n] \in \ncalS} \mathbbm{1} \left( |\Re \{\chi(l;a[\cdot])\}| > u \right) \nbbP(s[n] = a[n]) \notag\\
     &\geq \mathbbm{1} \left( |\Re \{\chi(l;a_0[\cdot])\}| > u \right) \nbbP(s[n] = a_0[n]) \notag \\
     &= \nbbP(s[n] = a_0[n])~(\mbox{for sufficiently small $u$}) \notag \\
     &= 2^{- m_s K} = \exp(-(m_{\rm s}\ln 2)  r N) \notag
 \end{align}
\end{IEEEproof}
\begin{nrem}[Implication of Theorems \ref{thm:main} and \ref{thm:lb}] 
\label{rem:corr_implication}
The combination of results from Theorems~\ref{thm:main} and \ref{thm:lb} implies that for linear codes, a faster order of decay of the auto-correlation function, in terms of the block length, is not possible. Importantly, the structure of a linear code (i.e., its generator/parity-check matrices) does not impact the order of the auto-correlation decay.
\end{nrem}

\begin{cor}[Asymptotically Ideal Auto-Correlation]
\label{cor:asymp}
% \footnote{Strictly speaking, the convergence is to a scaled $\delta$-function, with the scaling factor equal to the average symbol power, see (\ref{def:autocorr}).}
As $N \rightarrow \infty$, $|\Re\{\chi(l;s[\cdot])\}|$ converges in distribution to $\delta[l]$, i.e.,
\begin{align}
    \lim_{N \rightarrow \infty} |\Re \{\chi(l;s[\cdot]) \}| \xrightarrow{d}  \delta[l]. 
\end{align}
\begin{IEEEproof} For $l\neq0$ and $u>0$, we have from Theorem~\ref{thm:main}, 
  \begin{align}
     0 \leq \lim_{N \rightarrow \infty}  \nbbP \left( |\Re \{\chi(l;s[\cdot]) \}| > u \right) &\leq \lim_{N \rightarrow \infty} 2\exp(-O(rN)u^2) = 0 \notag\\
     \label{eq:conv_distrib}
    \therefore ~ \lim_{N \rightarrow \infty}  \nbbP \left( |\Re \{\chi(l;s[\cdot]) \}| > u \right) &= 0
 \end{align}
From (\ref{eq:conv_distrib}) and Definition~\ref{def:conv_d}, it follows that $|\Re\{\chi(l;s[\cdot])\}|$ converges in distribution to the deterministic $\delta[l]$, as $N \rightarrow \infty$. 
\end{IEEEproof}
\end{cor}

\subsection{Tail probability of $|\Re\{\varrho(l;s^{(i)}[\cdot],s^{(q)}[\cdot])\}|$}
\label{subsec:tail_prob_cc}
Assuming the symbols in $u^{(i)}[n]$ and $u^{(q)}[m]$ are mutually independent, we can derive the following bounds for $\Re\{\varrho(l;s^{(i)}[\cdot],s^{(q)}[\cdot])\}$, similar to Theorems~\ref{thm:main} and \ref{thm:lb}.
 
\begin{cor}[Cross-correlation Upper Bound]
\label{cor:main}
For $u > 0$, the tail probability of $|\Re\{\varrho(l;s^{(i)}[\cdot],s^{(q)}[\cdot])\}|$ satisfies the following upper bound:
 \begin{align}
\nbbP(|\Re\{\varrho(l;s^{(i)}[\cdot],s^{(q)}[\cdot])\}| > u) \leq 2\exp(-O(\max(r_i,r_q)N)u^2),
 \end{align}
as $K_i, K_q, N \rightarrow \infty$, $K_i/N=r_i$, and $K_q/N = r_q$. 
\end{cor}
\begin{IEEEproof}
Let $\tilde{K}:= \max(K_i-l, K_q)$. Similar to Definition~\ref{def:Ys}, $\Re\{\varrho(l;s^{(i)}[\cdot],s^{(q)}[\cdot])\}$ can be expressed as follows:
 \begin{align}
 \label{eq:tildeYs}
     \Re\{\varrho(l;s^{(i)}[\cdot],s^{(q)}[\cdot])\} &:= \frac{1}{N} \left(\tilde{Y}_1^{(l)} + \cdots + \tilde{Y}_{\tilde{M}_l}^{(l)}\right), \\
\mbox{where}~  \tilde{M}_l &:= \left\lceil \frac{N-l}{\tilde{K}} \right\rceil, \notag \\
\tilde{Y}_a^{(l)} &:= \Re\left\{\sum_{k' = 0}^{\tilde{K}-1} s^{(i)}[(a-1)\tilde{K}+l+k'] s^{(q)^*}[(a-1)\tilde{K}+k'] \right\}, ~a=1,\cdots,\tilde{M}_l-1 \notag \\
\tilde{Y}_{\tilde{M}_l}^{(l)} &:= \Re\left\{s^{(i)}\left[\left(\tilde{M}_l -1\right)\tilde{K}+l\right] s^{(q)^*}\left[ \left(\tilde{M}_l -1 \right)\tilde{K} \right] \right. \notag \\
&\hspace{10mm} \left. + s^{(i)}\left[\left(\tilde{M}_l -1\right)\tilde{K}+l+1\right] s^{(q)^*}\left[ \left(\tilde{M}_l -1 \right)\tilde{K}+1 \right] \right. \notag\\
&\hspace{10mm} \left. + \cdots + s^{(i)}[N-1] s^{(q)^*}[N-1-l] \right\}
 \end{align}
The remainder of the proof follows along the same lines as Theorem~\ref{thm:main}. 
\end{IEEEproof}

\begin{cor}[Cross-correlation Lower Bound]
\label{cor:lb}
For linear codes, the tail probability of $|\Re \{\varrho(l; s^{(i)}[\cdot], s^{(q)}[\cdot]))|$ satisfies the following lower bound for $l \ll N$ and \underline{sufficiently small} $u>0$:
  \begin{IEEEeqnarray}{rCl}
     \nbbP(|\Re\{\varrho(l;s^{(i)}[\cdot],s^{(q)}[\cdot])\}| > u) & \geq & \exp(-O((r_i + r_q) N)),
 \end{IEEEeqnarray}
where $K_i/N=r_i$ and $K_q/N=r_q$. 
\end{cor}
\begin{IEEEproof}
For linear codes, the joint probability of the all-zero codeword for both $s^{(i)}[n]$ and $s^{(q)}[n]$ equals $2^{-(m_{{\rm s}_i}K_i + m_{{\rm s}_q}K_q)}$. The remainder of the proof follows along the same lines as Theorem~\ref{thm:lb}.%It is easily seen from (\ref{eq:varrho}) that when all the symbols in $s^{(i)}[n]$ and $s^{(q)}[n]$ are identical -- e.g., the all-zero codeword for both signals -- then, $|\varrho(l;s^{(i)}[\cdot], s^{(q)}[\cdot])| > u$, for $l \ll N$ and sufficiently small $u$. The  
\end{IEEEproof}

\begin{nrem}[Implications of Corollary~\ref{cor:main} and \ref{cor:lb}]
\label{rem:crosscorr_implication}
Similar to Remark~\ref{rem:corr_implication}, the results from Corollaries~\ref{cor:main} and \ref{cor:lb} imply that for linear codes, a faster rate of decay of the cross-correlation function, in terms of the block length, is not possible.
\end{nrem}

\begin{cor}[Asymptotically Zero Cross-Correlation]
\label{cor:asymp_crosscorr}
As $N \rightarrow \infty$, $|\Re\{\varrho(l;s^{(i)}[\cdot],s^{(q)}[\cdot])\}|$ converges in distribution to 0, for any $l$, i.e.,
\begin{align}
     \lim_{N \rightarrow \infty} |\Re\{\varrho(l; s^{(i)}[\cdot], s^{(q)}[\cdot])\}|  \xrightarrow{d} 0, ~ \mbox{ any $l$}.
 \end{align}
\end{cor}
\begin{IEEEproof}
 Similar to Corollary~\ref{cor:asymp}.
\end{IEEEproof}

\begin{nrem}[Communications versus Sensing trade-off w.r.t favourable correlation properties]
\label{rem:decay_tradeoff}
Broadly, there are two contrasting mechanisms for obtaining signals with (nearly) favourable correlation properties: (a) through deterministic construction (e.g., m-sequences, Zadoff-Chu sequences etc.), or (b) through LLN, exploiting (information-bearing) randomness (e.g., c.c.s). Signals belonging to the first class typically have \underline{better} correlation properties relative to c.c.s (i.e., lower sidelobes for a given $N$), and therefore, have excellent sensing performance, but without any communication value\footnote{Some communication value can be embedded to the deterministic sensing signals by modulating them with information-bearing symbols \cite{Mao_et_al_2021}, but the data rates that can be achieved by this method are very low. On the other hand, the randomness from the information-bearing symbols contributes to reducing the sidelobe levels for such signals \cite{Eedara_et_al_2018}.} (e.g., data rate). At the other end of the spectrum, the information-bearing randomness in c.c.s offers high communication value at the expense of potentially larger sidelobes, that in turn, may restrict its effectiveness as sensing signals to specific scenarios (e.g., small-range applications, where the near-far effect is less prevalent).
\end{nrem}

\subsection{Tail Probability of $|\Re \{V^{(q \rightarrow i)}[l] \}|$}
\label{subsec:tail_prob_ofdm}
Similar to Corollaries~\ref{cor:main} and \ref{cor:lb}, we can derive the following bounds for $\Re \{V^{(q \rightarrow i)}[l]\}$, for mutually independent $u^{(i)}[n]$ and $u^{(q)}[n]$.

\begin{cor}[OFDM Interference Sidelobe Upper Bound]
\label{cor:idft_sidelobe}
 For $u > 0$, the tail probability of $\Re\{V^{(q \rightarrow i)}[l]\}~(l=0,\cdots,N-1)$ satisfies the following upper bound:
 \begin{align}
 \label{eq:thm_idft}
 \nbbP \left( |\Re \{V^{(q \rightarrow i)}[l] \}| > u \right) \leq 2\exp(-O(\min(r_i,r_q) N) u^2), 
 \end{align}
as $K_i, K_q, N \rightarrow \infty$, $K_i/N = r_i$, and $K_q/N=r_q$. 
\end{cor}
\begin{IEEEproof}
 Let $K_0:= \min(K_i,K_q)$. Similar to Definition~\ref{def:Ys}, $\Re\{V^{(q \rightarrow i)}[l]\}$ can be expressed as follows :
\begin{align}
    \Re\{V^{(q \rightarrow i)}[l]\} &:= \frac{1}{N} \left(T_1[l] + \cdots + T_{M_0}[l]\right), \notag\\
\label{eq:T_first_l}    
\mbox{where}~ M_0 &:= \left \lceil \frac{N}{K_0} \right\rceil, \notag \\
T_i[l] &:=  \Re\left\{ \sum_{k= (i-1)K_0}^{iK_0 - 1}  \frac{s^{(q)}[k]}{s^{(i)}[k]} \exp\left(j 2\pi \frac{l}{N} k \right) \right\}; ~i=1,\cdots,M_0-1, \\
\label{eq:T_last_l}
\mbox{and}~ T_{M_0}[l] &:= \Re\left\{ \sum_{k= (M_0-1)K_0}^{N - 1}  \frac{s^{(q)}[k]}{s^{(i)}[k]} \exp\left(j 2\pi \frac{l}{N} k \right) \right\}. 
\end{align}
The remainder of the proof follows along the same lines as Theorem~\ref{thm:main}.
\end{IEEEproof}

\begin{cor}[OFDM Interference Sidelobe Lower Bound]
\label{cor:ofdm_lb}
For linear codes, the tail probability of $\Re\{V^{(q \rightarrow i)}[0]\}$ satisfies the following lower bound for sufficiently small $u>0$:
  \begin{align}
     \nbbP(|\Re\{ V^{(q \rightarrow i)}[0] \}| > u) \geq \exp(-O((r_i + r_q) N)),
 \end{align}
where $K_i/N=r_i$ and $K_q/N=r_q$. 
\end{cor}
\begin{IEEEproof}
 From (\ref{eq:def_V}), we observe that when all the symbols in $s^{(i)}[n]$ and $s^{(q)}[n]$ are identical -- e.g., the all-zero codeword for both signals -- then, $|\Re\{ V^{(q \rightarrow i)}[0] \}| > u$, for sufficiently small $u$. The remainder of the proof follows along the same lines as Corollary~\ref{cor:lb} and Theorem~\ref{thm:lb}. 
\end{IEEEproof}

\begin{nrem}[Implications of Corollary~\ref{cor:idft_sidelobe} and \ref{cor:ofdm_lb}]
\label{rem:ofdm_bounds_implication}
Similar to Remarks~\ref{rem:corr_implication} and \ref{rem:crosscorr_implication}, the results from Corollaries~\ref{cor:idft_sidelobe} and \ref{cor:ofdm_lb} imply that for linear codes, a faster rate of decay of $V^{(q\rightarrow i}[l]$, in terms of the block length, is not possible.
\end{nrem}

\begin{cor}[Asymptotic Interference Suppression]
\label{cor:asymp_idft}
As $N \rightarrow \infty$, $|\Re\{V^{(q \rightarrow i)}[l]\}|$ converges in distribution to 0, for any $l$, i.e.,
 \begin{align}
     \lim_{N \rightarrow \infty} |\Re\{V^{(q \rightarrow i)}[l]\}|  \xrightarrow{d} 0, ~ \mbox{ any $l$}.
 \end{align}
\end{cor}
\begin{IEEEproof}
 Similar to Corollary~\ref{cor:asymp}.
\end{IEEEproof}

\begin{nrem}[Single-Carrier v/s OFDM (Common) Waveform for ISAC -- Impact of c.c.s on Sensing Performance]
The similarity in the asymptotic behavior of the the tail probabilities of $\chi(l;s[\cdot])~(l\neq 0)$, $\varrho(l; s^{(i)}[\cdot], s^{(q)}[\cdot])$ and $V^{(q \rightarrow i)}[l]$ -- i.e., the $\exp(-O(rN))$ decay -- implies that for any code rate, c.c.s are effective sensing signals that are robust to sensing interference at \underline{sufficiently large} block lengths, with negligible difference in performance based on whether they modulate a single-carrier or OFDM waveform. We investigate this claim in the following section.
\end{nrem}

\section{Simulation Results}
\label{sec:sim_res}
In this section, we first verify the bounds derived in Section~\ref{sec:corr}, before comparing the range-Doppler sensing performance of c.c.s in interference-limited operation with that of a benchmark interference-free FMCW waveform.

\subsection{Peak-to-Sidelobe Ratio and Sensing Interference Suppression of c.c.s}
\label{subsec:check}
The sidelobes\footnote{The sidelobes of $\chi(l;s[\cdot])$ have two main contributors: (a) the correlation properties of $s[n]$, and (b) timing errors. These two factors are independent, in the sense that even if $\chi(l;s[\cdot]) = \delta[l]$, sidelobes can still be present if there is a timing offset. In this paper, we focus on the sidelobes due to (a) only, assuming no timing errors. The sidelobes due to (b) can be minimized using a pulse shape with a fast roll-off (e.g., root-raised cosine or Gaussian pulses).} of $\chi(l;s[\cdot])$, can be measured using the \emph{peak-to-sidelobe ratio} (PSLR) metric, defined as follows:
\begin{align}
\label{eq:pslr}
    {\rm PSLR~(dB)} &:= -20\log_{10} \left(\max_{l \neq 0} |\chi(l;s[\cdot])|\right),
\end{align}
where a large value signifies a closer approximation to $\delta[l]$. Similarly, replacing $\chi(l;s[\cdot])$ with  $\varrho(l;s^{(i)}[\cdot], s^{(q)}[\cdot])$ and $V^{(q \rightarrow i)}[l]$ in (\ref{eq:pslr}) yields a metric that measures the extent of interference suppression provided by c.c.s for single-carrier and OFDM waveforms, respectively.

\begin{figure}
\centering
\begin{subfigure}{0.5\textwidth}
    \centering
    \includegraphics[scale = 0.55]{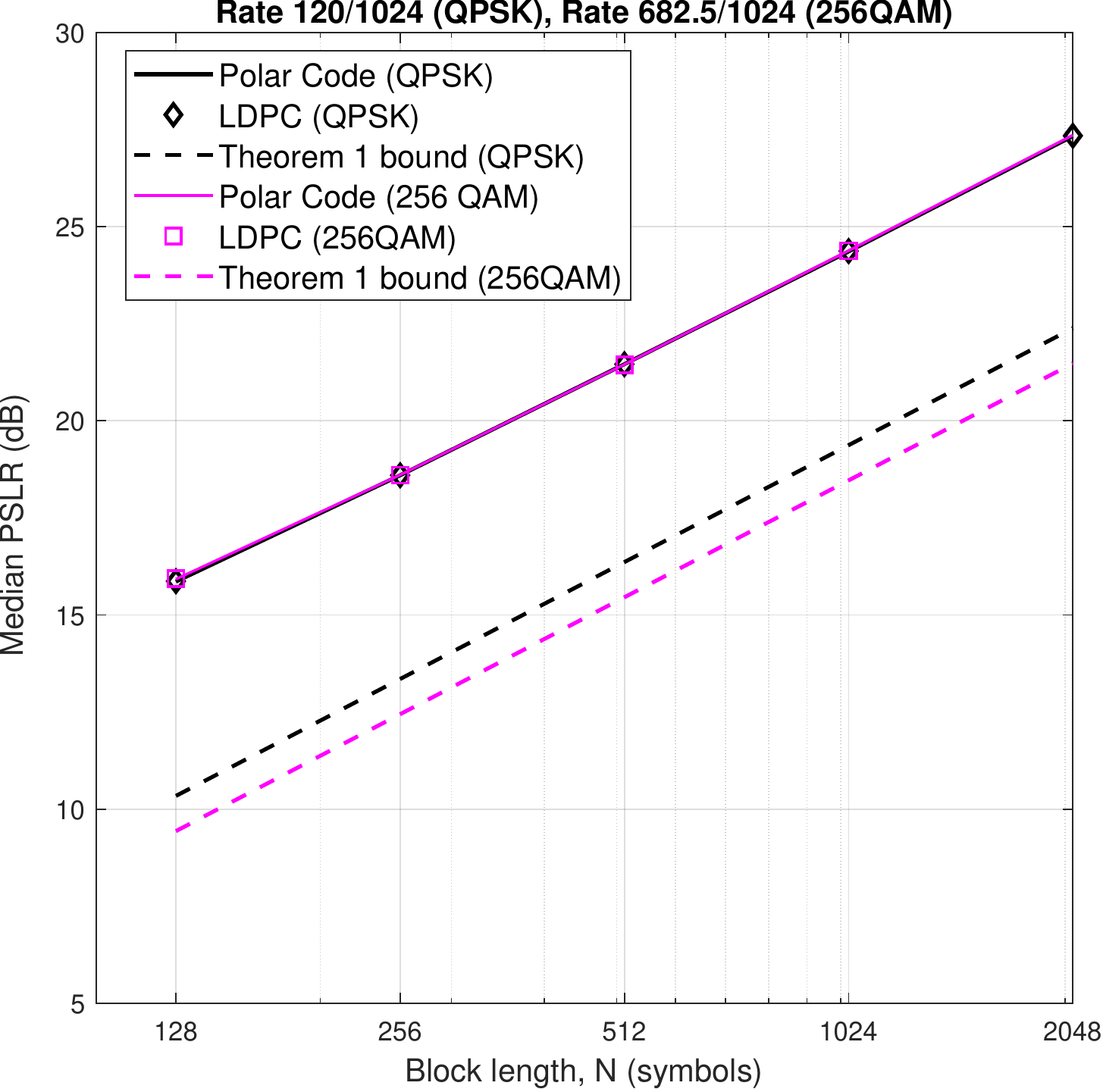} 
    \caption{PSLR of c.c.s}
\end{subfigure}%
~
\begin{subfigure}{0.5\textwidth}
    \centering
    \includegraphics[scale = 0.55]{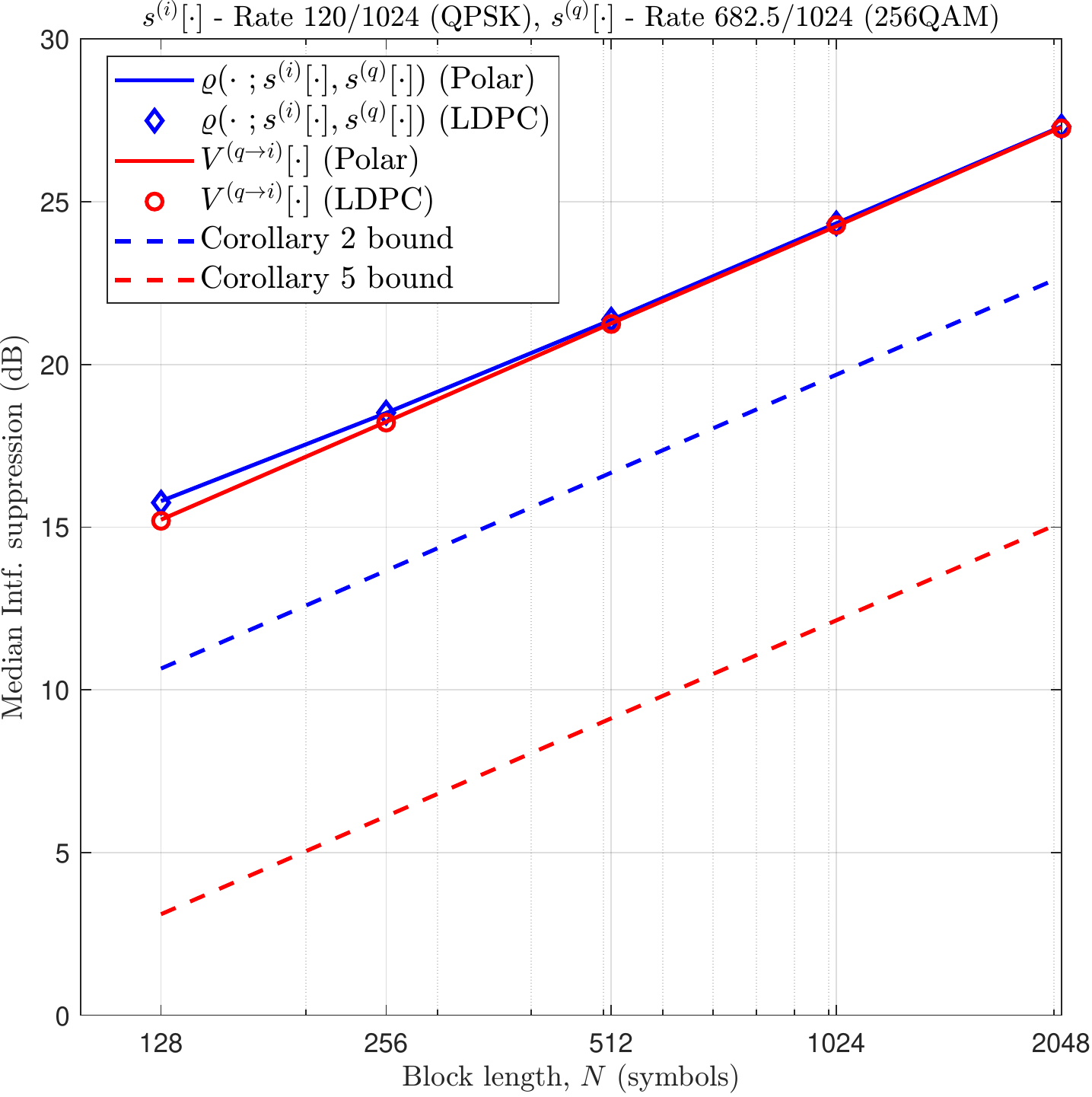}     \caption{Interference Suppression of c.c.s}
\end{subfigure}
\caption{The median PSLR and mutual interference suppression provided by c.c.s increases by approximately $3{\rm dB}$ when the block length is doubled, mirroring the bounds in Theorem~\ref{thm:main}, Corollary~\ref{cor:main} and Corollary~\ref{cor:idft_sidelobe}.}%However, the PSLR of c.c.s is lower than that of an FMCW waveform at large block lengths. Hence, the latter is better suited for sensing applications featuring large target ranges, where the near-far effect is more likely to be prevalent. In contrast, c.c.s can be effective sensing signals for small-range applications.}
\label{fig:pslr}
\end{figure}

Figs.~\ref{fig:pslr}a and \ref{fig:pslr}b respectively plot the median PSLR and interference suppression (obtained over 10000 simulation instances) as a function of $N$, for LDPC and Polar coded signals\footnote{Unlike the analysis in Section~\ref{sec:corr}, we have not imposed systematic encoding (Assumption~\ref{assump:sys_encoding}) in our simulations.} at two rates -- $120/1024$ and $682.5/1024$ -- corresponding to the smallest code rates for which QPSK and 256QAM are used in 5G NR, respectively \cite[Table~5.1.3.1-2]{3gpp_mcsspecs}. We make the following remarks, based on the plotted curves:

\begin{nrem}[$3{\rm dB}$ increase in (median) PSLR and Interference Suppression with a doubling of the block length]
\label{rem:3dB}
Theorem~\ref{thm:main}, Corollary~\ref{cor:main} and Corollary~\ref{cor:idft_sidelobe} each provide lower bounds\footnote{A lower bound is obtained for a suitable scaling factor of the $O(rN)$ term. We have assumed a value of $1/2b^2$, where $b$ is defined in Lemma~\ref{lem:charY1} and depends on the symbol constellation.}, which suggests that the median PSLR and interference suppression provided by c.c.s should eventually increase by at least $3{\rm dB}$ when the block length is doubled; for the values of $N$ considered, we observe that this holds true, as the curves corresponding to c.c.s are nearly parallel to those obtained from the bounds.
\end{nrem}

\begin{nrem}[Impact of Code structure]
\label{rem:coding_impact}
% The code structure is not important
The convergence behaviour of $\chi(l; s[\cdot])~(l\neq 0)$, $\varrho(l;s^{(i)}[\cdot],s^{(q)}[\cdot])$ and $V^{(q \rightarrow i)}[l]$ is governed by LLN and the ``extent of i.i.d-ness" among the codeword symbols (Remark~\ref{rem:iid_extent}), which does not depend on the code structure (i.e., Polar or LDPC) for fixed $r$, $N$ and modulation scheme. Hence, the code structure has negligible impact on the median PSLR and interference suppression provided by c.c.s.
\end{nrem}

\begin{nrem}[The Effectiveness of Interleaving]
A key component of the c.c.s model in Fig.~\ref{fig:coding_setup} is the interleaver, and in Fig.~\ref{fig:pslr}, the codeword bits are interleaved according to a uniformly distributed random permutation. To underline the effectiveness of interleaving, Fig.~\ref{fig:interleaver} compares the median PSLR with and without interleaving for Polar coded c.c.s, where we observe the following:
\begin{itemize}
    \item From the bound in Theorem~\ref{thm:main}, we observe that the median PSLR (in dB) is proportional to $10\log_{10}(rN)$. Hence, for code rates \underline{sufficiently close} to one, the block length has a much greater influence on the median PSLR. This is reflected in the curves for uninterleaved c.c.s, where (a) the median PSLR increases with $r$ for fixed $N$, but the performance at $r = 682.5/1024$ practically coincides with that of an uncoded signal (i.e., $r = 1$); and (b) the difference in the median PSLR between low and high rate c.c.s diminishes with $N$. 
    
    \item Regardless of the modulation and coding scheme, the interleaved c.c.s has the same median PSLR as that of an uncoded signal. This suggests that the interleaved c.c.s -- comprising mutually uncorrelated symbols -- is \underline{ergodic}, despite exhibiting statistical dependence over a block. The seeming ergodicity of the interleaved c.c.s may also explain why the bounds derived in Section~\ref{sec:corr} are conservative in Fig.~\ref{fig:pslr} -- the bounds are based on the extent of i.i.d-ness within c.c.s (see Remark~\ref{rem:iid_extent}) and not on their ergodic properties.
\end{itemize}
In summary, interleaving considerably improves the PSLR (and similarly, the interference suppression) of c.c.s, especially for small code rates and block lengths.
\end{nrem}

\begin{nrem}[Effective Sensing Interference Suppression]
Higher interference suppression at larger block lengths is a noteworthy feature of c.c.s that is shared with other well-known deterministic sensing signals (e.g., m-sequences), but is not a universal feature among the latter, in general; for instance, a pair of FMCW chirp signals with different chirp slopes gives rise to a distinctive interference pattern that is not straightforward to eliminate \cite{Uysal_Sanka_2018, Kim_Mun_Lee_2018}. However, we note that this robustness is restricted to multi-user interference, and not any form of adversarial interference, like jamming.
\end{nrem}

\begin{nrem}[Near-far effect]
\label{rem:near_far}
The PSLR of a sensing waveform is a measure of its robustness to the near-far effect. To illustrate this, consider two identical targets -- one near and one far from the radar -- and suppose $N = 1024$, for which the median PSLR is $\approx 25{\rm dB}$. Due to $R^{-4}$ attenuation, the radar return from the far target is $> 25~{\rm dB}$ weaker than that of the near one, when the former's range exceeds the latter's by a factor $> 4.2$. Thus, c.c.s are effective sensing signals only up to a certain maximum range that depends on the block length. We explore this in more detail in Section~\ref{subsec:nearfar}.
\end{nrem}

%\begin{nrem}[c.c.s as sensing signals for small-range applications]
%\label{rem:range}
%The PSLR of the FMCW waveform (see for its expression) is higher than that of c.c.s at larger block lengths. The main drawback associated with a low PSLR is the possibility of the \emph{near-far effect}.  

%As a result, the FMCW waveform is better suited for sensing applications featuring large target ranges.  
%\end{nrem}

% Why is PSLR similar in distribution to cross-PSLR

%With the exception of $l=0$, $\chi_s[l]$ and $\varrho_{1,2}[l]$ are identically distributed, with the same extent of i.i.d-ness (Remark~\ref{rem:iid_extent}); hence, it is intuitive that the median interference suppression in Fig.~\ref{fig:cross_pslr} coincides with the median PSLR in Fig.~\ref{fig:pslr}. 

% Consequently, analogous to Remark~\ref{rem:3dB}, a doubling of the block length increases the median interference suppression by approximately $3{\rm dB}$, which 

\begin{figure}
    \centering
    \includegraphics[scale = 0.7]{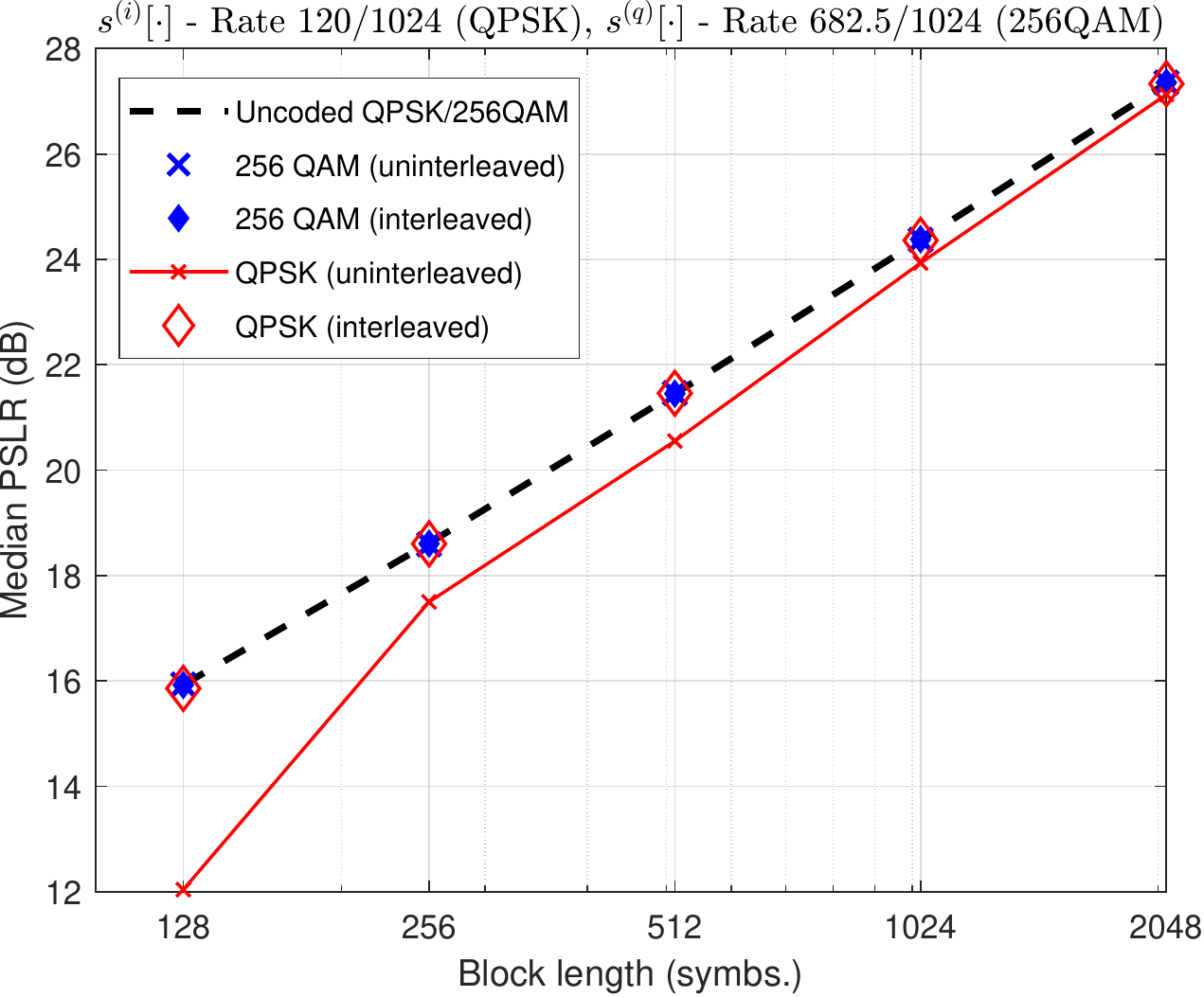}
    \caption{Interleaving improves the PSLR of c.c.s, especially for small code rates and block lengths.}
    \label{fig:interleaver}
\end{figure}

\subsection{Sensing Performance of c.c.s and the Near-Far Effect}
\label{subsec:nearfar}
In this subsection, we explore the sensing performance of c.c.s in an interference-limited scenario featuring the near-far effect, by considering the radar scene in Fig.~\ref{fig:setup}b with TX~$1$ as the desired radar, TX~$2$ as the interfering radar, and two (point) targets (i.e., $M_r = N_{\rm tar} = 2$). We model this scene in terms of the notation in Section~\ref{subsec:sensing_model} below (see Table~\ref{tab:sim} for a full list of simulation parameters): \begin{itemize}
    \item[(i)] With respect to TX~$1$, suppose the near target is at $d_1 = 2{\rm m}$ and moving at $v_1 = 10{\rm mph}$, while the far target is at $d_2 = 4{\rm m}$ and moving at $v_2 = 15{\rm mph}$ in the same direction as the near target. Assuming a signal bandwidth of  1~{\rm GHz}, a carrier frequency of $140~{\rm GHz}$, and $M = N = 1024$, the targets' (range, Doppler) bins are $(n_1^{(1)}, m_1^{(1)}) = (14, 516)$ and $(n_2^{(1)}, m_2^{(1)}) = (27, 518)$. Note that the Doppler bins can vary for a different choice of $M$.

    \item[(ii)] We assume that the interference experienced at TX~$1$ is dominated by the signal received along the direct path, TX~$2 \rightarrow$ TX~$1$, corresponding to $l=0$ in the RHS of (\ref{eq:y_intf_pair}). Hence, for simplicity, we consider only this component and assume $(n_0^{(2 \rightarrow 1)},m_0^{(2 \rightarrow 1)}) = (29,518)$, which corresponds to TX~$2$ situated at $d_{\rm intf} = 4.3{\rm m}$ away from TX~$1$ and moving at $v_{\rm intf} = 15{\rm mph}$ -- i.e., close to and moving as fast as the far target in (i) above.

    \item[(iii)] For the amplitudes of the desired radar returns, we assume $\alpha_1^{(1)} = 1$ and $\alpha_2^{(1)} = 10^{-12/10}$, resulting in a $24{\rm dB}$ difference in the return energy between the near and far targets. Of this, $12{\rm dB}$ is due to $R^{-4}$ attenuation [i.e., $\propto (d_1/d_2)^4$], while the rest is accounted for by assuming a weaker radar cross section for the far target. 
    
    \item[(iv)] Unlike the radar returns, the interference signal experiences $R^{-2}$ attenuation, and therefore, $\frac{|\alpha_0^{(2 \rightarrow 1)}|^2}{|\alpha_1|^2} \propto \frac{d_1^4}{d_{\rm intf}^2}$. Hence, we assume $\alpha_0^{(2 \rightarrow 1)} = \sqrt{10^{11/10}}$ -- based on a unit radar cross section for the near target -- resulting in an SIR of $-11{\rm dB}$ w.r.t the near target. With these parameter values, a couple of key question are whether (a) the strong signal from the \emph{interference bin} is suppressed, and (b) the weak return from the far target bin is visible in the range-Doppler maps? We explore these in Fig.~\ref{fig:rd_map}.

    \item[(iv)] The c.c.s blocks, $s^{(1)}[n,m]$ and $s^{(2)}[n,m]$, are obtained from a rate $120/1024$ Polar code, followed by QPSK modulation. The corresponding message blocks, $u^{(1)}[n,m]$ and $u^{(2)}[n,m]$, are mutually independent. Finally, the SNR w.r.t to the near target is $0{\rm dB}$.
    %    \item[(v)]  
\end{itemize}

\begin{table}[t]
    \centering
    \begin{tabular}{|c|c|c|c|}
    \hline
     Bandwidth, $B$ & $1~{\rm GHz}$ & Target Ranges $(d_1, d_2)$ & $2{\rm m}$, $4{\rm m}$\\
     Range resolution, $\Delta r$ & $0.15{\rm m}$ & Target Range bins $(n_1^{(1)},n_2^{(1)})$ & $14$, $27$ \\
     Block length, $N$ & $1024$ (symbols) & Target speeds $(v_1, v_2)$ & $10{\rm mph}$, $15{\rm mph}$  \\
     Largest range bin, $n_{\rm max}$ & $32$ & Target Doppler bins $(m_1^{(1)},m_2^{(1)})$ & $516$, $518$\\

     Maximum detectable range & $4.65{\rm m}$ & Target return gains $(\alpha_1^{(1)}, \alpha_2^{(1)})$ & $1$, $10^{-12/10}$ \\
     \hline
     Carrier frequency, $f_c$ & $140~{\rm GHz}$ & Intf.~range ($d_{\rm intf}$) & $4.3{\rm m}$ \\
     No. of blocks, $M$ & $1024$ & Intf.~range bin ($n_0^{(2\rightarrow 1)}$) & $29$\\
     Velocity resolution, $\Delta v$ & $2.28{\rm mph}$ & Intf. speed ($v_{\rm intf}$) & $15{\rm mph}$\\
     \cline{1-2}
     Coding & Polar Code ($r = 120/1024$) & Intf. Doppler bin ($m_0^{(2\rightarrow 1)}$) & 518\\
     Modulation & QPSK & Intf. gain ($\alpha_0^{(2\rightarrow 1)}$) & $\sqrt{10^{11/10}}$ \\
     SNR (w.r.t near target) & $0{\rm dB}$ & & \\
     SIR (w.r.t near target) & $-11{\rm dB}$ & & \\
    \hline
    \end{tabular}
    \caption{List of simulation parameters for the near-far target scenario in Section~\ref{subsec:nearfar}.}
    \label{tab:sim}
\end{table}

The presence/absence of a target in (range, Doppler) bin $(l,\nu)$ is decided using a threshold rule, as follows: 
\begin{align}
\label{eq:det_rule}
    H_1 (H_0)&:~ \mbox{Target present (absent)} \notag \\
    |R_w^{(1)}[l,\nu]| &\underset{H_0}{\overset{H_1}{\gtrless}}  \eta,~ w \in \{{\rm sc, OFDM}\}.
\end{align}
where $R_w^{(1)}[l,\nu]$ denotes the range-Doppler map at TX~$1$ for $w\in \{\rm sc, OFDM\}$, defined in (\ref{eq:rd_map}) and (\ref{eq:ofdm_rd_map}). For a given threshold, $\eta$, the sensing performance of c.c.s is characterized by the detection and false alarm probabilities -- denoted by $P^w_d(\eta)$ and $P^w_f (\eta)$, respectively -- and defined as follows:
\begin{IEEEeqnarray}{rCl}
    \label{eq:Pd}
     P^w_d (\eta) &:= & \underbrace{\nbbP\left(\left|R_w^{(1)}[n_1^{(1)},m_1^{(1)}]\right| > \eta;~\left|R_w^{(1)}[n_2^{(1)},m_2^{(1)}]\right| > \eta \right)}_\text{Both targets detected} \notag \\
     & & + \underbrace{\frac{1}{2}\nbbP\left(\left|R_w^{(1)}[n_1^{(1)},m_1^{(1)}]\right| > \eta;~\left|R_w^{(1)}[n_2^{(1)},m_2^{(1)}]\right| \leq \eta \right)}_\text{Only near target detected (hence, the 1/2 scaling factor)} \notag \\
     & & + \underbrace{\frac{1}{2}\nbbP\left(\left|R_w^{(1)}[n_1^{(1)},m_1^{(1)}]\right| \leq \eta;~\left|R_w^{(1)}[n_2^{(1)},m_2^{(1)}]\right| > \eta \right)}_\text{Only far target detected (hence, the 1/2 scaling factor)}, \\
    \label{eq:Pf}
    P_f^w (\eta) &:= &\nbbP\left( \left|R_w^{(1)}(l,\nu)\right| > \eta, ~\mbox{where}~ (l,\nu)\notin \{(n_1^{(1)},m_1^{(1)}), (n_2^{(1)},m_2^{(1)})\} \right).
\end{IEEEeqnarray}
\begin{figure}[t]
    \centering
    \begin{subfigure}{0.99\textwidth}
     \centering
        \includegraphics[scale = 0.75]{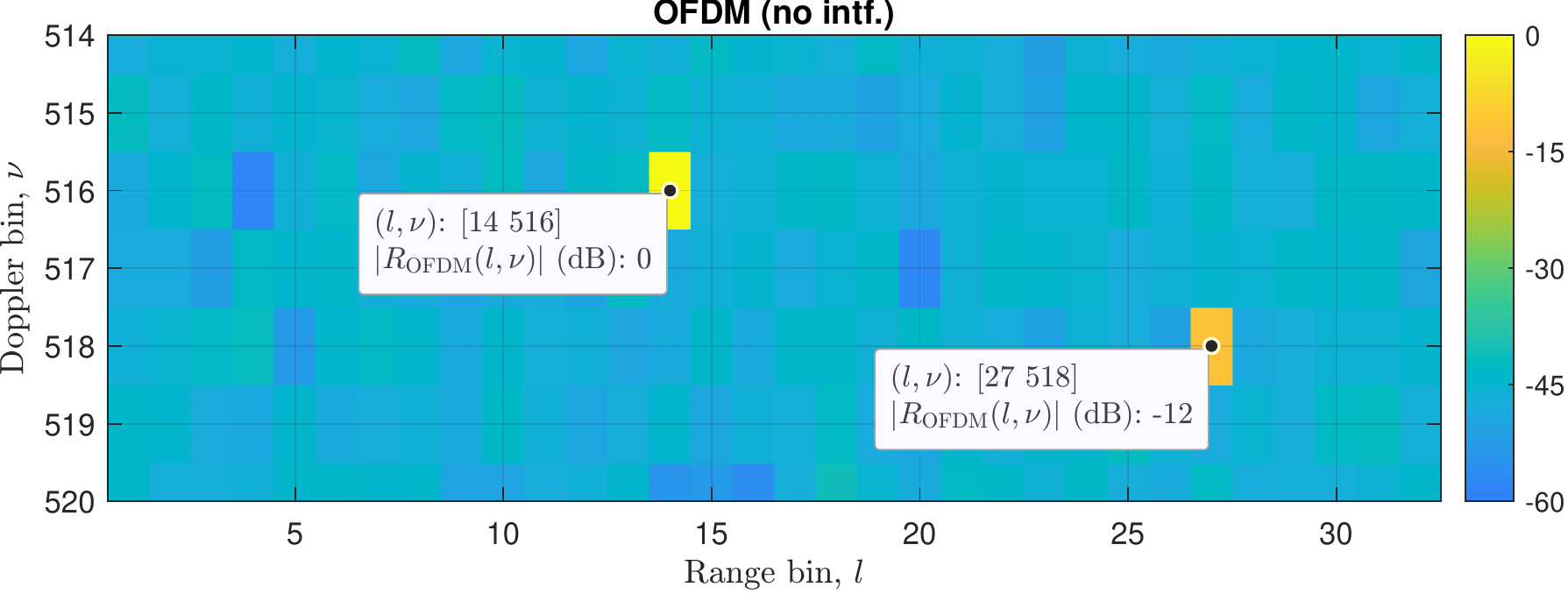}
    \end{subfigure}%
    \\
    \begin{subfigure}{0.99\textwidth}
     \centering
        \includegraphics[scale = 0.75]{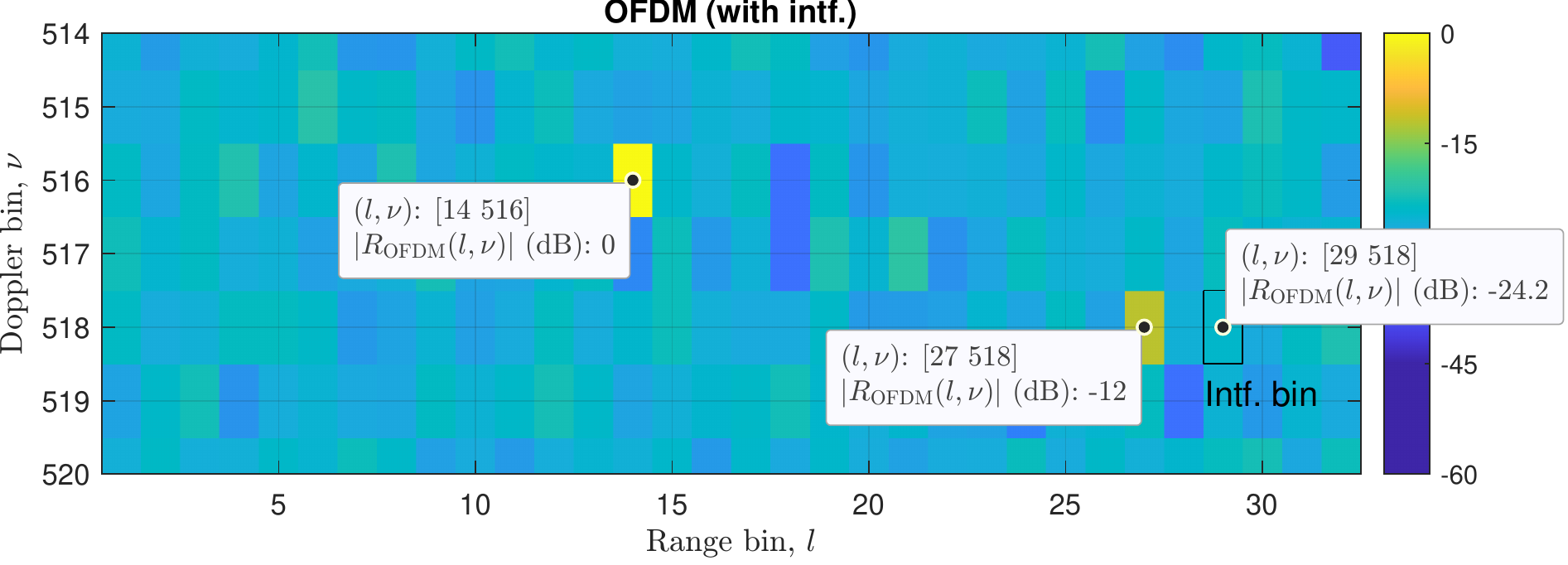}
    \end{subfigure}%
    \\
    \begin{subfigure}{0.99\textwidth}
     \centering
        \includegraphics[scale = 0.75]{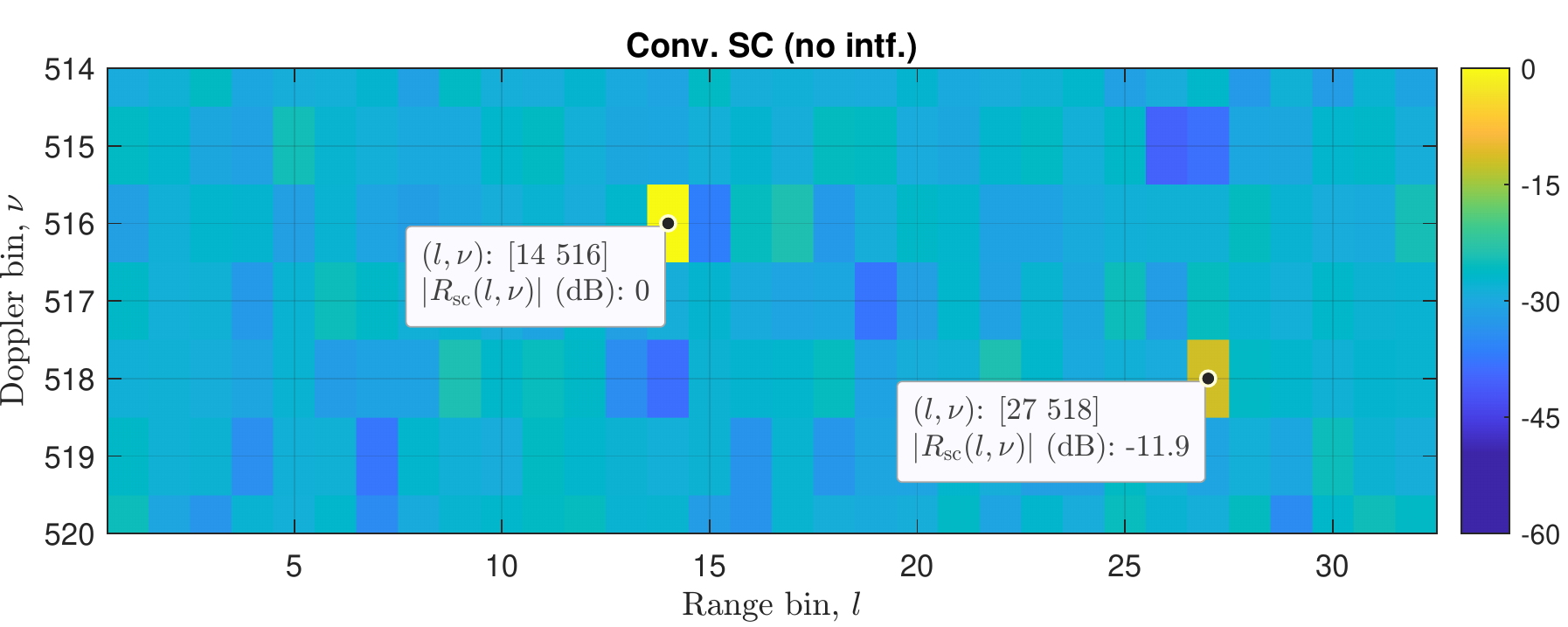}
    \end{subfigure}%
    \\
    \begin{subfigure}{0.99\textwidth}
    \centering
        \includegraphics[scale = 0.75]{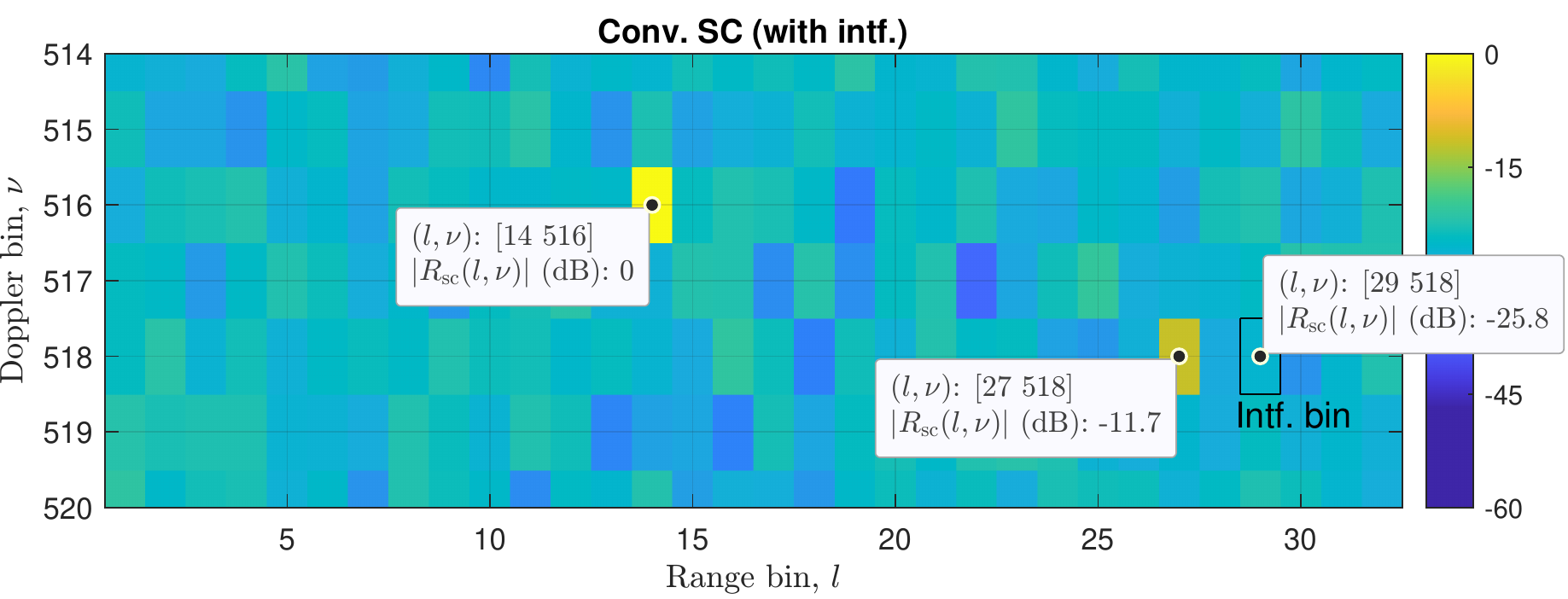}
    \end{subfigure}%
    \caption{Continued in the next page}
    \end{figure}
    \begin{figure}[t]
    \ContinuedFloat
    \begin{subfigure}{0.99\textwidth}
    \centering
        \includegraphics[scale = 0.75]{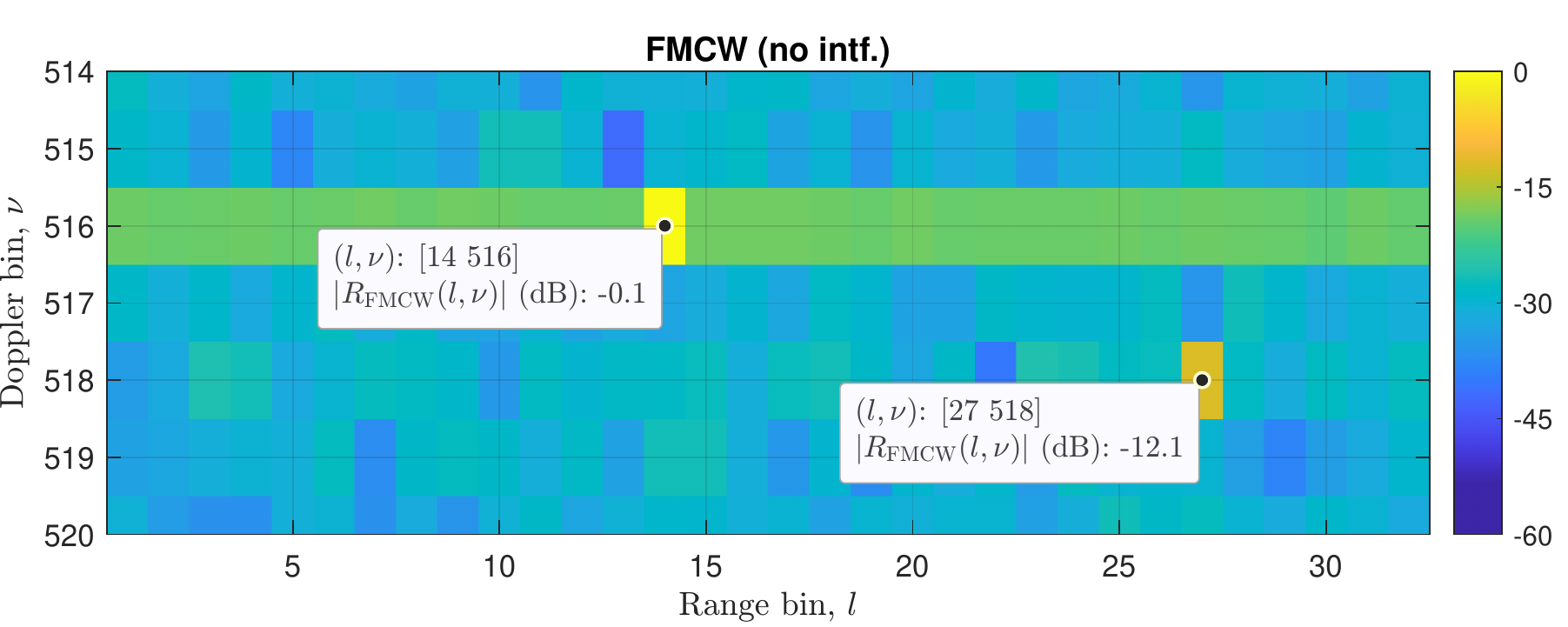}
    \end{subfigure}%
    \caption{For the simulation setup in Table~\ref{tab:sim}, the magnitude of the range-Doppler maps at the target bin locations -- $(14,516)$ and $(27,518)$ -- are similar for both FMCW and c.c.s waveforms, even at $-11{\rm dB}$ SIR for the latter.}
    \label{fig:rd_map}
\end{figure}

Fig.~\ref{fig:rd_map} shows the range-Doppler maps for a given realization of c.c.s and FMCW waveforms. We observe that:
\begin{itemize}
    \item For $N = 1024$, the median PSLR [$\approx 25{\rm dB}$ in Fig.~\ref{fig:pslr}(a)] is nearly at par with the difference in return strengths between the near and far targets. Despite this, the far target is clearly ``visible" because the targets are situated in different Doppler bins, which provides nearly $30{\rm dB}$ of additional sidelobe suppression for $N=1024$ through the Doppler DFT.
    
    \item For $N=1024$, the nearly $25{\rm dB}$ median interference suppression in Fig.~\ref{fig:pslr}b seems insufficient, at first glance, to detect the weaker target in the face of strong interference from a nearby bin. However, from (\ref{eq:pslr}), we see that the interference suppression is defined in terms of the maximum sidelobe level \emph{across all} bins, which can be a conservative estimate of the sidelobe levels at a specific interference bin. In this case, we see that the strong interference signal from a nearby bin is adequately suppressed and the weaker target is clearly visible. However, the essence of Remark~\ref{rem:near_far} -- i.e., a limit on the maximum target range imposed by the sidelobes of c.c.s -- remains valid.
    
    \item The magnitude of the c.c.s range-Doppler maps at the target bin locations is nearly the same as that for the FMCW waveform -- both with and without interference and for both single-carrier and OFDM waveforms. This is consistent with the results from Section~\ref{sec:corr} that c.c.s yield asymptotically ideal range-Doppler maps (i.e., $\delta[\cdot]$ spikes at the target bins) at large block lengths, despite the presence of sensing interference.
    
    \item The large sidelobes across the range bins in the FMCW range-Doppler map are a consequence of rectangular windowing assumed for the range FFT. Lower sidelobe levels can be achieved through a better choice of windowing function, which we do not pursue as it is beyond the scope of this paper.
\end{itemize}

% Explanation of step-like behavior
\begin{figure}[b]
        \centering
        \includegraphics[scale = 0.85]{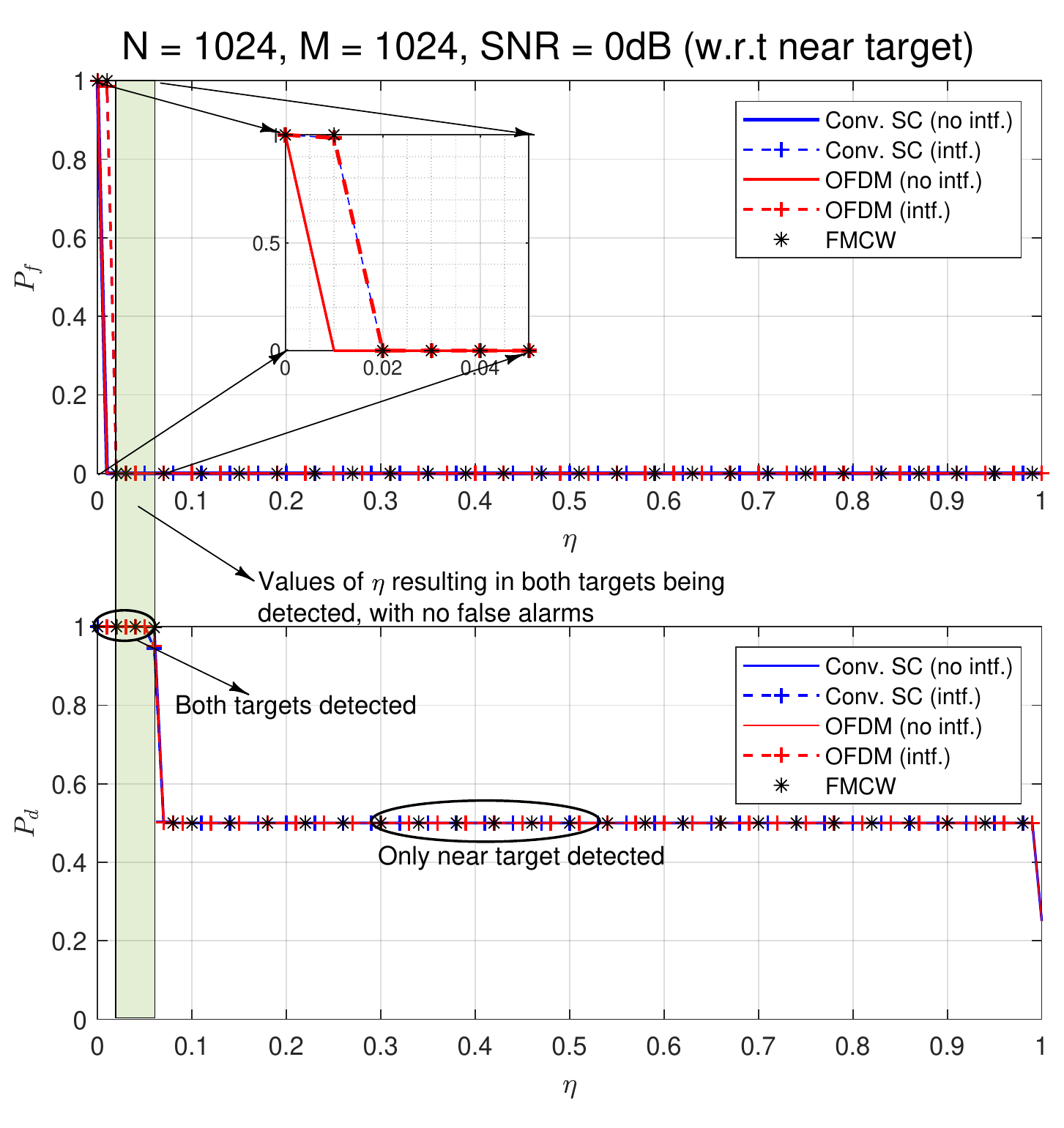}
        \caption{Due to the similarity of the range-Doppler maps in Fig.~\ref{fig:rd_map}, the curves corresponding to the detection and false-alarm probabilities for all the waveforms also exhibit considerable overlap.}
        \label{fig:PfPd}
\end{figure}

Fig.~\ref{fig:PfPd} compares the detection performance of c.c.s, where we observe the following: 
\begin{itemize}
    \item For a threshold-based detection rule of the form given by (\ref{eq:det_rule}), the detection probability exhibits a step-like behavior with increasing $\eta$, where for small $\eta$, both targets are detected; followed by a regime where only the near target is detected; before none of the targets is detected. %The false-alarm probability, on the other hand, is non-increasing in $\eta$, as expected.
    
    \item For a small set of thresholds, the FMCW waveform has a higher false-alarm probability than c.c.s without interference (see inset), due to the large sidelobes induced by rectangular windowing, as discussed w.r.t Fig.~\ref{fig:rd_map}. Thus, window functions that result in lower sidelobes would also cause the false-alarm probability curve in Fig.~\ref{fig:PfPd} to shift leftward.
    
    \item Despite strong interference ($-11{\rm dB}$ SIR), the detection performance of c.c.s matches that of an interference-free FMCW waveform, and is only marginally worse -- in terms of the behavior of the false-alarm probability -- than c.c.s without interference.
    
    \item The \emph{sweet-spot} threshold(s) for a given waveform are those that result in both targets being detected, without any false-alarms. From the previous bullet point, it follows that the sweet spot regions for c.c.s and FMCW mostly coincide. 
% In situations where the near-far effect is unlikely to occur, the sensing performance of the c.c.s waveform can even be better than that of the FMCW waveform.
    
%    \item %For $N = 1024$, the matched filter provides a post-processing SNR gain of nearly $30{\rm dB}$; hence, even the return from the weaker target will be $6{\rm dB}$ above the noise floor. On the other hand, the 
% Thus, the detection of the far target is limited by the PSLR. Although the median PSLR ($\approx 25{\rm dB}$ in Fig.~\ref{fig:pslr}(a)) is similar in magnitude

\end{itemize}
We conclude this section by noting that the results from Figs.~\ref{fig:pslr} through \ref{fig:PfPd} corroborate the bounds derived in Section~\ref{sec:corr}, thereby demonstrating that c.c.s are effective sensing signals even in the presence of strong sensing interference, with negligible difference in performance based on whether they modulate a single-carrier or OFDM waveform.

\section{Concluding Remarks}
\label{sec:conc_rems}
In this paper, we explored the sensing potential of c.c.s in interference-limited ISAC scenarios featuring both multi-user communications and sensing interference, with the latter dominating the former. We started by identifying -- based on whether the waveform was single-carrier or OFDM -- functions of c.c.s that could adversely affect its sensing performance through large sidelobes. We then derived upper bounds for the tail probabilities of the sidelobe levels that decayed exponentially in terms of the code rate-block length product, which suggested that c.c.s were effective sensing signals that were robust to sensing interference at sufficiently large block lengths for any fixed code rate, with negligible difference in performance based on whether they modulate a single-carrier or OFDM waveform. The latter implication was verified through simulations, where we observed that the sensing performance of c.c.s -- in terms of detection and false-alarm probabilities -- was at par with that of the interference-free FMCW waveform for both single-carrier and OFDM waveforms, even at an SIR of $-11~{\rm dB}$. Thus, our results imply that (a) a common ISAC waveform (either single- or multi-carrier, but) largely comprising coded data symbols is an effective sensing signal at large block lengths, and (b) in multi-user ISAC scenarios with such a waveform, sensing interference management essentially \emph{takes care of itself} for monostatic radars, and is relatively simpler than communications interference management. These are highly favourable outcomes in terms of the evolution of existing wireless networks to support sensing applications while also maximizing the communications spectral efficiency, as the sensing functionality does not impose additional constraints on the available resources, either in the form of needing more reference signals or orthogonal resource allocation across users.

% whose tail probabilities needed to decay rapidly for them to be effective sensing signals robust to multi-user sensing interference. In Section~\ref{sec:corr}, we derived upper bounds for these tail probabilities that 

\appendices
\section{Repetition Codes: Interleaving mitigates correlation across codeword bits}
\label{app:interleave}
Let $\nbc = [c_0 ~\cdots~ c_{\gamma K-1}]=[\underbrace{b_0~\cdots~b_{K-1}~\cdots~b_0~\cdots~b_{K-1}}_\text{$\gamma$ times}]$ denote the codeword corresponding to the i.i.d. Bernoulli($1/2)$ message vector $\nbb = [b_0 ~\cdots b_{K-1}]$ for a rate $1/\gamma$ repetition code, where $\gamma \in \{1,2,\cdots\}$. Then, for $i \neq j$ and $j > i$, the correlation coefficient of $c_i$ and $c_j$ -- denoted by $\rho(c_i,c_j)$ -- is given by:
\begin{align}
\label{eq:bitlevelcorr}
    \rho(c_i, c_j) = \begin{cases}
                        1, ~ |j-i| \mod K = 0 \\
                        0,~ \mbox{else},
                       \end{cases}
\end{align}
which captures the fact that codeword bits that are separated by a multiple of $K$ are fully correlated.

Let $\tilde{\nbc} = \pi(\nbc)$ denote the interleaved codeword, where $\pi(.)$ is a permutation of bit locations such that: (a) $\pi(i) = i$, for $i=0,\cdots,K-1$ (i.e., systematic encoding from Assumption~\ref{assump:sys_encoding}), and (b) $[\pi(K),~ \cdots ~, \pi(\gamma K-1)]$ is a uniformly distributed random permutation of $\{K, K+1 \cdots,\gamma K - 1\}$. Then, for $i\neq j$ and $j>i$
\begin{align}
\label{eq:interleave_cond1}
 \rho(\tilde{c}_i, \tilde{c}_j) &= \nbbP(|\pi^{-1}(i) - \pi^{-1}(j)| \mod K = 0) \notag \\
 &= \begin{cases}
  0, \hspace{3mm} i,j \in \{0,\cdots,K-1\} \\
  \frac{\gamma - 1}{(\gamma-1)K}, \hspace{3mm} i\in \{0,\cdots,K-1\}, j \in \{K,\cdots,\gamma K - 1\} \\
  \frac{\gamma - 2}{(\gamma-1)K-1}, \hspace{3mm} i,j \in \{K, \cdots, \gamma K - 1 \}.
 \end{cases}
\end{align}
From (\ref{eq:interleave_cond1}), we see that $\rho(\tilde{c}_i \tilde{c}_j) \rightarrow 0$ as $K \rightarrow \infty$. Thus, the interleaved codeword bits are asymptotically uncorrelated.

\bibliographystyle{IEEEtran}
\bibliography{IEEEabrv,mmWaveRadCom_bib}
\end{document}